\documentclass[]{mn2e}
\usepackage[english]{babel}
\usepackage[T1]{fontenc}
\usepackage{amsmath}
\usepackage{amssymb}
\usepackage{booktabs}
\usepackage{graphicx,subfig}
\usepackage{color}
\usepackage{epstopdf}
\usepackage{comment}
\usepackage{caption}

\newcommand*{\Ledd}{L_{\rm Edd}}
\newcommand*{\calR}{{\cal R}}
\newcommand*{\calM}{{\cal M}}

\newcommand*{\Mdotedd}{\dot M_{\rm Edd}}
\newcommand*{\Mdotb}{\dot M_{\rm B}}
\newcommand*{\Mdotbe}{\dot M_{\rm e}}
\newcommand*{\Mdotbes}{\dot M_{\rm es}}

\newcommand*{\Mdott}{\dot M_{\rm t}}
\newcommand*{\Mdotacc}{\dot M_{\rm acc}}

\newcommand*{\lambdat}{\lambda_{\rm t}}

\newcommand*{\Lambdacr}{\Lambda_{\rm cr}}
\newcommand*{\lambdacr}{\lambda_{\rm cr}}
\newcommand*{\lambdaes}{\lambda_{\rm es}}

\newcommand*{\gmin}{g_{\rm min}}
\newcommand*{\fmin}{f_{\rm min}}
\newcommand*{\xmin}{x_{\rm min}}
\newcommand*{\xz}{x_0}

\newcommand*{\Mbh}{M_{\rm BH}}
\newcommand*{\Mg}{M_{\rm g}}
\newcommand*{\rhog}{\rho_{\rm g}}

\newcommand*{\cs}{c_{\rm s}}
\newcommand*{\rb}{r_{\rm B}}
\newcommand*{\rbe}{r_{\rm e}}
\newcommand*{\rg}{r_{\rm g}}

\newcommand*{\phit}{\phi_{\rm t}}
\newcommand*{\phig}{\phi_{\rm g}}

\newcommand*{\rhoinf}{\rho_{\infty}}
\newcommand*{\pinf}{p_{\infty}}
\newcommand*{\csinf}{c_{\infty}}
\newcommand*{\Tinf}{T_{\infty}}
\newcommand*{\rhotil}{\tilde\rho}
\newcommand*{\mpr}{m_{\rm p}}

\title[Bondi accretion in galaxies]
{Bondi accretion in early-type galaxies}
 \author[V. Korol, L. Ciotti and S. Pellegrini]{Valeriya Korol$^{1,2,}$\thanks{E-mail:
korol@strw.leidenuniv.nl}, Luca Ciotti$^1$ and Silvia Pellegrini$^1$
\\$^1$Department of Physics and Astronomy, University of Bologna, viale Berti Pichat
6/2, 40127 Bologna, Italy
\\$^2$Leiden Observatory, Leiden University, PO Box 9513, 2300 RA Leiden, the Netherlands}

\date{February 13, 2016}

\pagerange{\pageref{firstpage}--\pageref{lastpage}} \pubyear{2016}

\begin{document}
\maketitle
\label{firstpage}

\begin{abstract}
  Accretion onto central massive black holes in galaxies is often
  modelled with the Bondi solution.  In this paper we study a
  generalization of the classical Bondi accretion theory, considering
  the additional effects of the gravitational potential of the host
  galaxy, and of electron scattering in the optically thin limit. We
  provide a general analysis of the bias in the estimates of the Bondi
  radius and mass accretion rate, when adopting as fiducial values for
  the density and temperature at infinity the values of these
  quantities measured at finite distance from the central black
  hole. We also give general formulae to compute the correction terms
  of the critical accretion parameter in relevant asymptotic
  regimes. A full analytical discussion is presented in the case of an
  Hernquist galaxy, when the problem reduces to the discussion of a
  cubic equation, therefore allowing for more than one critical point
  in the accretion structure.  The results are useful for
  observational works (especially in the case of low-luminosity
  systems), as well as for numerical simulations, where accretion
  rates are usually defined in terms of the gas properties near the
  black hole.
\end{abstract}


\section{Introduction} 

The Bondi solution of accretion on a point mass (Bondi 1952), due to
its inherent simplicity, is a standard tool for the interpretation of
observations of the accretion phenomenon, and the starting point for
the development of recipes for the mass accretion rate, to be adopted
for example in semi-analytical models and numerical simulations that
lack the resolution to study gas transport down to parsec scale.  In
cosmological simulations and semi-analytical models of the early
growth of massive black holes (hereafter MBHs), and of the
co-evolution of MBHs and their host galaxies, the Bondi accretion rate
is used to link the mass supply to the accretion disks surrounding
MBHs with the density and temperature of their environment (e.g.,
Fabian \& Rees 1995, Volonteri \& Rees 2005, Di Matteo et al. 2005,
Hopkins et al. 2006, Booth \& Schaye 2009, Park \& Ricotti 2011,
Wyithe \& Loeb 2012, Hirschmann et al. 2014, Inayoshi et al. 2015,
DeGraf et al. 2015, Curtis \& Sijacki 2015).  Another important
application of the classical Bondi model is to estimate the mass
accretion rate on MBHs at the center of galaxies, by using observed
values of the gas density and temperature in the vicinity of the MBH
(e.g., Loewenstein et al. 2001; Baganoff et al. 2003, Pellegrini 2005,
2010; Allen et al. 2006; Rafferty et al. 2006; McNamara et al. 2011;
Wong et al. 2014; Russell et al. 2015).  In this view, one assumes
that these values represent the true boundary conditions (i.e., at
infinity) for the Bondi problem (see Quataert \& Narayan 2000).

However, it is recognized that even the knowledge of the true boundary
conditions would not be enough for a proper treatment of mass
accretion on MBHs at the center of galaxies: first, because the MBH is
not isolated, being at the bottom of the host galaxy potential well;
second, because the radiation emitted by the inflowing material
interacts with the material itself, with the consequent establishment
of unsteady accretion (for luminosities of the order of $10^{-2}\Ledd$
or greater, where $\Ledd$ is the Eddington luminosity; e.g., Cowie et
al. 1978); and finally, because the flow also gets mass and energy
from the inputs due to stellar evolution (e.g., Ciotti et al. 1991).
When the last two of the above circumstances are important, Bondi
accretion cannot be applied; during phases of moderate accretion,
instaed, the problem can be considered almost steady, so that Bondi
accretion can be considered a first, reliable approximation of the
real situation.

In this paper we present a quantification of the bias on the estimates
of the Bondi radius and mass accretion rate, that is introduced when
adopting as boundary values for the density and temperature those at
arbitrary but finite distances from the MBHs.  First we derive the
exact formulae for this bias, in case of radiation pressure due to
electron scattering, and of the additional gravitational potential of
a galaxy; we also derive the asymptotic expansion of these formulae
close to the MBH.  These formulae contain a critical accretion
parameter, that in general can be determined only numerically. Then,
we present a technique to obtain the analytical expressions for the
critical accretion parameter, in some special cases.  A full
analytical discussion of the critical points is presented for a
Hernquist galaxy model.  We finally solve numerically the Bondi
problem for a MBH at the center of a galaxy, and including the effect
of radiation pressure due to electron scattering, and compare the
numerical results with the analytical ones.

The paper is organized as follows.  In Section 2 we recall the main
properties of the classical Bondi solution, and we present a
preliminary analysis of the mass accretion bias introduced by
considering as boundary values for the density and temperature those
at points along the solution at finite distance from the MBH.  In
Section 3 we add to the Bondi solution, in a self-consistent way, the
effect of electron scattering, and in Section 4 we consider the full
case of the Bondi solution in presence of radiation feedback and a
galaxy potential. In Section 5 the particular case of a Hernquist
galaxy is presented, building numerically the accretion solution, and
also providing a full analytical discussion of the problem.  For all
these cases we derive the formulae that allow to recover the true
accretion rate from fiducial estimates of the Bondi radius and
accretion rate obtained by assuming classical Bondi accretion.  The
main conclusions are summarized in Section 6. Finally, three
Appendixes contain technical details and relevant formulae useful in
analytical and numerical studies.

\section{The classical Bondi model} 
\label{sec:class}

As the present investigation builds on the classical Bondi (1952)
accretion model, it is useful to recall its main properties.  The
classical Bondi theory describes spherically-symmetric, steady
accretion of a spatially infinite gas distribution onto an isolated
central mass, in our case a MBH, of mass $\Mbh$. The self-gravity,
angular momentum and viscosity of the accreting gas, as well as
magnetic fields and feedback phenomena, are not considered.  The gas
is taken to be perfect, and subject to polytropic transformations;
thus its pressure ($p$) and density ($\rho$) are related by:
\begin{equation} 
p = {k_{\rm B} \rho T\over \mu\mpr} = {\pinf\over\rhoinf^{\gamma}} \rho^{\gamma},
\end{equation}
where $1 \le \gamma \le 5/3$ is the polytropic index, $\mpr$ is the
proton mass, $\mu$ is the mean molecular weight, $k_{\rm B}$ is the
Boltzmann constant, and $\pinf$ and $\rhoinf$ are respectively the gas
pressure and the density at infinity.  The polytropic gas sound speed
is
\begin{equation}
\cs^2=\gamma {p\over\rho}.
\end{equation}
Note that $\gamma$ is not necessarily the adiabatic index, so that in
the Bondi theory the entropy of the gas can change along the radial
streamlines (in fact, polytropic transformations
have a constant specific heat, e.g., Chandrasekhar 1939), and in
principle, once the solution is known, one could compute the
heat balance of each fluid element as it moves
toward the MBH\footnote{For a polytropic transformation of index
  $\gamma$, adiabatic index $\gamma_{ad}$, and specific heat at
  constant volume $c_V$, the molar specific heat is $c=c_V
  (\gamma_{ad}-\gamma)/(1-\gamma)$. Therefore, when
  $1<\gamma<\gamma_{ad}$, a fluid element loses energy as it moves inward
  and heats.}.

In spherical symmetry the time-independent continuity equation is:
\begin{equation} 
4 \pi r^2 \rho(r) v(r)= \Mdotb,
\end{equation}
where $v(r)$ is the gas radial velocity, and $\Mdotb$ is the
time-independent accretion rate on the MBH.  The Bernoulli
equation, with the appropriate boundary conditions at infinity,
becomes:
\begin{equation}
{v(r)^2\over 2} + \Delta h(r) - {G\Mbh\over r} = 0,
\end{equation}
where, from eq. (1) and $\gamma>1$, $ \Delta h$ is given by:
\begin{equation} 
\Delta h\equiv \int_{\pinf}^p {dp\over \rho} =
{\csinf^2\over\gamma-1} 
\left[\left({\rho\over\rhoinf} \right)^{\gamma-1} -1\right],
\end{equation}
where $\csinf$ is the sound speed of the gas at infinity.  In the
isothermal case, $\gamma=1$ and $\Delta h = \csinf^2 \ln
(\rho/\rhoinf)$.  In the case of an adiabatic transformation, $h$ is
the enthalpy per unit mass, while it is just proportional to it for a
generic polytropic transformation.

A scale length of fundamental importance for the problem, the
so-called Bondi radius, is naturally defined as\footnote{Sometimes in
  the literature a factor of $2$ appears in the numerator of the
  definition of the Bondi radius.}
\begin{equation} 
\rb \equiv {G\Mbh\over\csinf^2},
\end{equation}
and eqs. (3)-(4) are then recast in dimensionless form by introducing
the normalized quantities:
\begin{equation}
x\equiv {r\over\rb},\quad
\rhotil\equiv {\rho\over\rhoinf},\quad
\tilde{\cs} \equiv {\cs\over\csinf}=\rhotil^{\gamma-1\over 2}, 
\end{equation}
and the Mach number $\calM=v/\cs$.
For $\gamma>1$, eqs. (3)-(4) then become
\begin{equation}
\begin{cases}
\displaystyle{x^2 \calM \rhotil^{\gamma+1\over 2}=\lambda,}
\\ \\
\displaystyle{
{\calM^2 {\tilde\cs}^2\over 2} + 
{\rhotil^{\gamma -1}\over \gamma -1} = {1\over x} +{1\over \gamma-1},}
\end{cases}
\end{equation}
where 
\begin{equation} 
\lambda \equiv {\Mdotb\over 4\pi\rb^2\rhoinf\csinf}
\end{equation}
is the dimensionless accretion parameter: once known, it fixes
the accretion rate for assigned  $\Mbh$
and boundary conditions for the accreting gas. Note how from
eqs. (7)-(8) it follows that all physical quantities can be
expressed in terms of the radial profile of the Mach number.  By elimination
of $\rhotil$ in eq. (8), the Bondi problem reduces to the solution of
the equation
\begin{equation} 
	g(\calM) = \Lambda f(x), \qquad \Lambda \equiv
       \lambda^{2 (1-\gamma)\over \gamma +1},
\end{equation}
where 
\begin{equation} 
\begin{cases}
\displaystyle{
	g(\calM) = \calM^{2 (1-\gamma)\over \gamma+1} 
                          \left( {\calM^2\over 2} + {1\over\gamma-1 }\right),}
\\ \\
\displaystyle{
	f(x)=x^{4 (\gamma-1)\over\gamma +1}\left({1\over x}
          + {1\over \gamma-1}\right).}
\end{cases}
\end{equation}

As well known, $\Lambda$ cannot be chosen arbitrarily: for $1 < \gamma
\le 5/3$, both $g(\calM)$ and $f(x)$ have a minimum (that we indicate
with $\gmin$ and $\fmin$, respectively), thus to satisfy eq. (10)
$\forall x > 0$ requires that $\gmin \le \Lambda \fmin$, i.e., that
$\Lambda \ge \Lambdacr\equiv\gmin/\fmin$.  Equation (10) then
implies
\begin{equation}
\lambda \le \lambdacr \equiv\left({\fmin\over\gmin}\right)^{\gamma+1\over 2(\gamma-1)}.
\end{equation}
It is easy to show that, for $\gamma >1$:
\begin{equation} 
\begin{cases}
\displaystyle{\calM_{\rm min}=1, \qquad \gmin={\gamma+1\over 2(\gamma-1)},}
\\ \\
\displaystyle{\xmin={5-3\gamma\over 4},\quad\fmin={\gamma+1\over
  4(\gamma-1)}\left({4\over 5-3\gamma}\right)^{5-3\gamma\over\gamma+1},}
\end{cases}
\end{equation}
so that for the classical Bondi problem one has: 
\begin{equation}
\lambdacr ={1\over 4}\left({2\over 5-3\gamma}\right)^{5-3\gamma\over 2(\gamma-1)}.
\end{equation}
Note that $\lambdacr= e^{3/2}/4$ for $\gamma \to 1^+$, and
$\lambdacr=1/4$ for $\gamma \to {5/3}^-$.

In the isothermal case, the analogous of eq. (10) is:
\begin{equation}
g(\calM)+\ln{\lambda}=f(x),
\end{equation}
where now
\begin{equation}
\begin{cases} 
\displaystyle{g(\calM)={\calM^2\over 2}-\ln\calM ,}
\\ \\
\displaystyle{f(x)={1\over x}+2\ln x}.
\end{cases}
\end{equation}
Solutions exist only for $\gmin+\ln\lambda \le \fmin$, i.e., for
$\lambda \le \lambdacr = e^{\fmin-\gmin}$.  Simple algebra shows that
\begin{equation}
\begin{cases}
\displaystyle{\calM_{\rm min}=1, \qquad \gmin={1\over 2},}
\\ \\
\displaystyle{\xmin={1\over 2}, \qquad \fmin= 2-2\ln 2,}
\end{cases}
\end{equation}
so that  $\lambdacr$ in the isothermal case coincides with
the limit of eq. (14) for $\gamma\to 1^+$. 

In practice, to solve the Bondi problem means to obtain the radial
profile of $\calM(x)$, for given $\Lambda \ge \Lambdacr$.
Unfortunately, eqs. (10) and (15) do not have an explicit solution,
and must be solved numerically.  We do not enter in the details of the
solutions (see, e.g., Bondi 1952; Frank, King \& Raine 1992; Krolik
1998) and, as common in similar studies, we restrict to the critical
case $\lambda=\lambdacr$; in this case, $x_{\rm min}$ is also the
sonic radius. Among the two critical solutions, we consider that with
increasing Mach number approaching the center. In the following, the
function $f(x)$ in eqs. (11) and (16) is generalized by considering
the effect of radiation pressure due to electron scattering, and the
additional gravitational field of the host galaxy. In those more
general cases, we provide the formulae for the true mass accretion
rate, together with its estimates obtained using values of density and
temperature at any radius $r$ along the (new) solution, while assuming
classical Bondi accretion. The critical solutions for each case are
constructed with the aid of a numerical code built on purpose: we
first determined numerically the position and the value of the
absolute minimum of $f$ (Sects. 4 and 5), and, after having determined
$\lambdacr$, we evaluated the solution over the whole radial range.

\subsection{Mass accretion bias: concepts}  
\label{sec:biasclass}

Here we introduce the general procedure that will be considered in the
next Sections to estimate the Bondi radius and mass accretion rate,
for the basic case of the classical Bondi solution.  To keep the
notation simple, in the following we use the symbol $\lambda$ to
indicate the critical value $\lambdacr$.  Later we will use the same
procedure after having included in the problem the effects of
radiation pressure due to electron scattering (Sect. 3), and of the
gravitational potential of the host galaxy (Sects. 4 and 5).

For assigned values of $\rhoinf$, $\Tinf$, $\gamma$ and $\Mbh$, the
classical Bondi accretion rate is given by eq. (9):
\begin{equation} 
	\Mdotb = 4 \pi \rb^2 \lambda \rhoinf\csinf.
\end{equation}

In practice, when dealing with observations or numerical
simulations, one inserts in eq. (18) the values of $\rho $ and $T$ at
a finite distance $r$ from the MBH, and considers them as ``proxies''
for $\rhoinf$ and $\Tinf$. This procedure gives an {\it estimated}
value of the Bondi radius (that we call $\rbe$) and mass
accretion rate (that we call $\Mdotbe$). Here we investigate how much
these $\rbe$ and $\Mdotbe$  depart from the true values $\rb$
and $\Mdotb$, as a function of $r$, under the assumption that the
Bondi solution holds at all radii.

The fiducial Bondi radius and mass accretion rate
are then defined as:
\begin{equation} 
\rbe (r)\equiv {G\Mbh\over\cs^2(r)},\quad
\Mdotbe(r) \equiv 4 \pi \rbe^2(r)  \lambda \rho(r)\cs (r).
\end{equation}
In particular, $\rbe$ can be conveniently normalized to $\rb$ as:
\begin{equation} 
{\rbe(x)\over\rb} = {\tilde\cs (x)}^{-2} =\rhotil(x)^{1-\gamma} = 
\left({x^2 \calM\over\lambda}\right)^{2(\gamma-1)\over \gamma+1},
\end{equation}
and, from eqs. (18)-(19), the ratio $\Mdotbe/\Mdotb$ can be expressed in 
terms of $\rbe/\rb$, independently of the boundary conditions 
$\rhoinf$ and $\Tinf$, as:
\begin{equation} 
{\Mdotbe(x)\over \Mdotb} = 
\left[{\rb\over\rbe(x)}\right]^{5-3\gamma\over 2(\gamma-1)}. 
\end{equation}
Obviously,
for $x\to \infty$, one has $\rbe \to \rb$, by definition of
$\rb$ and $\rbe$ [eqs. (6) and (19)]; or, equivalently, because $\rhotil
\to 1$; in turn, one also has that $\Mdotbe \to\Mdotb$.
Note also how, for $\gamma >1$, $\rbe$ is always smaller than $\rb$, since
the gas sound speed increases inward; for $\gamma=1$, the
sound speed is constant, and then $\rbe=\rb$, independently of the
distance from the center.

\begin{figure}
        \centering
        \subfloat{\includegraphics[height=0.45\textwidth, width=0.48\textwidth]{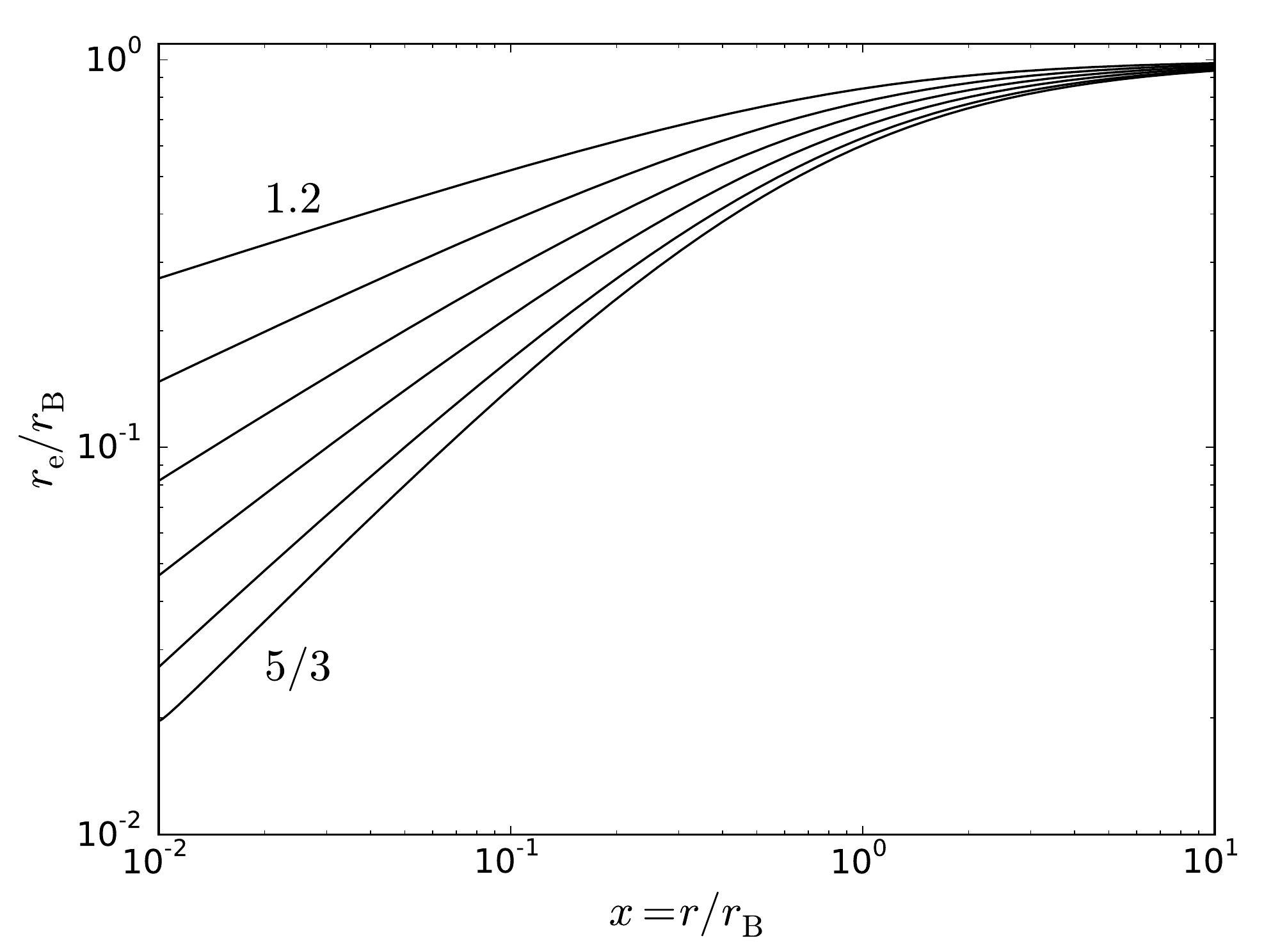}} \hfill
	\subfloat{\includegraphics[height=0.45\textwidth, width=0.48\textwidth]{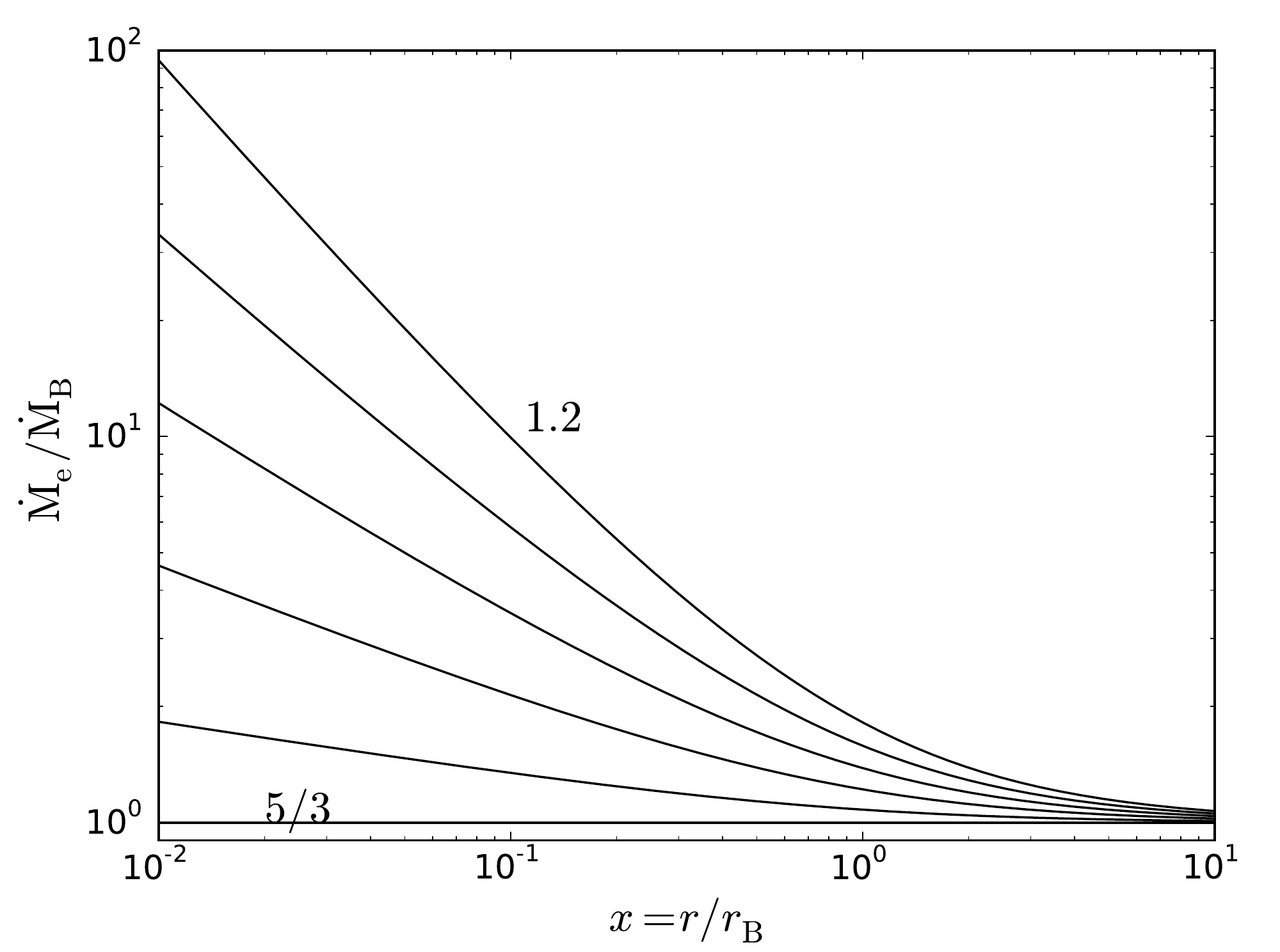}} \hfill
	\caption{Classical Bondi accretion model. {\it Upper panel:}
          estimated value of the Bondi radius $\rbe$ [eq. (20)], obtained from $T$ measured at a
          distance $r$ from the MBH, as a function of $r$.
	 {\it Lower panel:} estimated accretion rate
          $\Mdotbe$ in units of the true accretion rate $\Mdotb$ [eq. (21)], as a function of $r$. In both panels the polytropic index
          $\gamma$ is $1.2, 1.3, 1.4, 1.5, 1.6, 5/3$.}
       \label{fig:BIAS}
\end{figure}

For the mass accretion rate, in the monoatomic adiabatic case 
$(\gamma=5/3)$,
${\Mdotbe}(x)={\Mdotb}$, independently of the distance from the
center, i.e., there is no bias in the estimated mass accretion
rate. For $\gamma=1$, instead, using eq. (20), the bias is
just given by ${\rhotil(x)} $ at the radius $r$ where the measure is
taken, i.e., $\Mdotbe(x)=\rhotil(x)\Mdotb$.

A more quantitative insight in the behavior of eqs. (20)-(21) is
derived from the asymptotic expansion of $\calM (x)$ near the center,
which is obtained after computing the expansion of $f(x)$ for $x \to
0$, and of $g$ for $\calM \to \infty$, in eqs. (11) and (16). For
$1\leq\gamma\leq 5/3$, near the center $\calM \propto
x^{-{5-3\gamma\over 4}}$, and the leading terms of eqs. (20)-(21)
read:
\begin{equation} 
{\rbe(x)\over \rb} \sim
\left({\sqrt{2}\over\lambda}\right)^{\gamma-1}
x^{3(\gamma-1)\over 2},\quad  x \to 0^+,
\end{equation}
\begin{equation}
{\Mdotbe(x)\over\Mdotb} \sim 
\left({\lambda\over\sqrt{2}}\right)^{5-3\gamma\over 2}
x^{-{3(5-3\gamma)\over 4}}, \quad x \to 0^+. 
\end{equation}
For $\gamma >1$, eq. (22) confirms that $\rbe/\rb$ decreases as
$r$ approaches the center. For $\gamma=5/3$
($\lambda=1/4$), there is an interesting result: $\rbe\sim 2^{5/3}r$,
i.e., independently of the position $r$ at which the temperature to
derive $\rbe$ is taken, it is always concluded that the fiducial Bondi
radius is placed at a larger radius ($\rbe >r$), and by the same
factor.

For what is concerning the mass accretion rate $\Mdotbe $, from
eq. (23) one has that $\Mdotbe(x)/\Mdotb\propto x^{-3/2}$, if $\gamma
=1$; thus, $\Mdotbe$ significantly overestimates $\Mdotb$ for $x \to
0$, more than for any other larger $\gamma $. Note that in eq. (22) it
is possible to express $\rbe/\rb$ in terms of $r/\rbe$, and then in
eq. (21) to obtain $\Mdotb$ in terms of $\Mdotbe$ and $r/\rbe$. In
this way, the true $\Mdotb$ can be recovered from densities and
temperatures taken at some (small) distance $r$ from the center, as
$\Mdotb\sim (\sqrt{2}/\lambda)(r/\rbe)^{3/2}\Mdotbe (r)$. This
represents a useful result for observational and numerical studies,
affected by instrumental or grid resolution.

The properties above, together with the trend of $\rbe(x)/\rb$ with
$x$ for different $\gamma$, are illustrated by Fig. \ref{fig:BIAS},
that uses the numerical results of our code.  The figure shows how
$\rbe$ is smaller than $\rb$, and how this underestimate increases
with $r$ decreasing, and for increasing $\gamma$.  The lower panel of
Fig. \ref{fig:BIAS} shows the trend of $\Mdotbe(x)/\Mdotb$ with $x$,
for different $\gamma$; one sees that the use in eq. (19) of $\rho(r)$
instead of $\rhoinf$, of $\cs(r)$ instead of $\csinf$, and of $\rbe$
instead of $\rb$, leads to an overestimate of the true accretion rate
$\Mdotb$ (except for $\gamma=5/3$).  For $r<\rb$, the overestimate of
$\Mdotb$ is significant.  The numerical results in
Fig. \ref{fig:BIAS}, for $x\la 0.1$, are in excellent agreement with
those provided by the analytical asymptotic expansions near the center
in eqs. (22)-(23) (not shown in this figure for clarity; but see the
following Fig. 2).

\section{Adding the effects of electron scattering} 
\label{sec:escatt}

\begin{figure}
        \centering 
        \subfloat{\includegraphics[height=0.45\textwidth, width=0.48\textwidth]{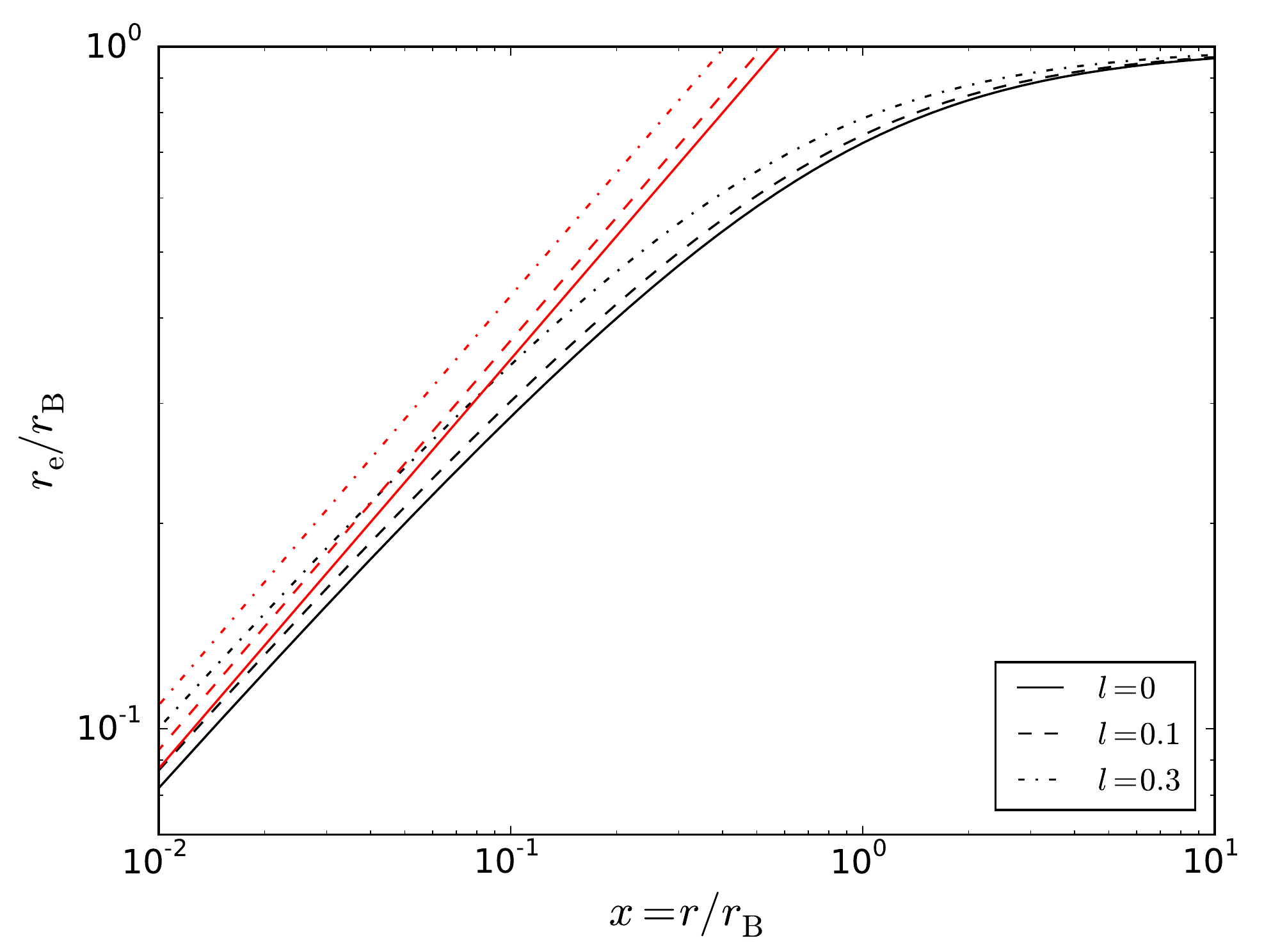}} \hfill 
	\subfloat{\includegraphics[height=0.45\textwidth, width=0.48\textwidth]{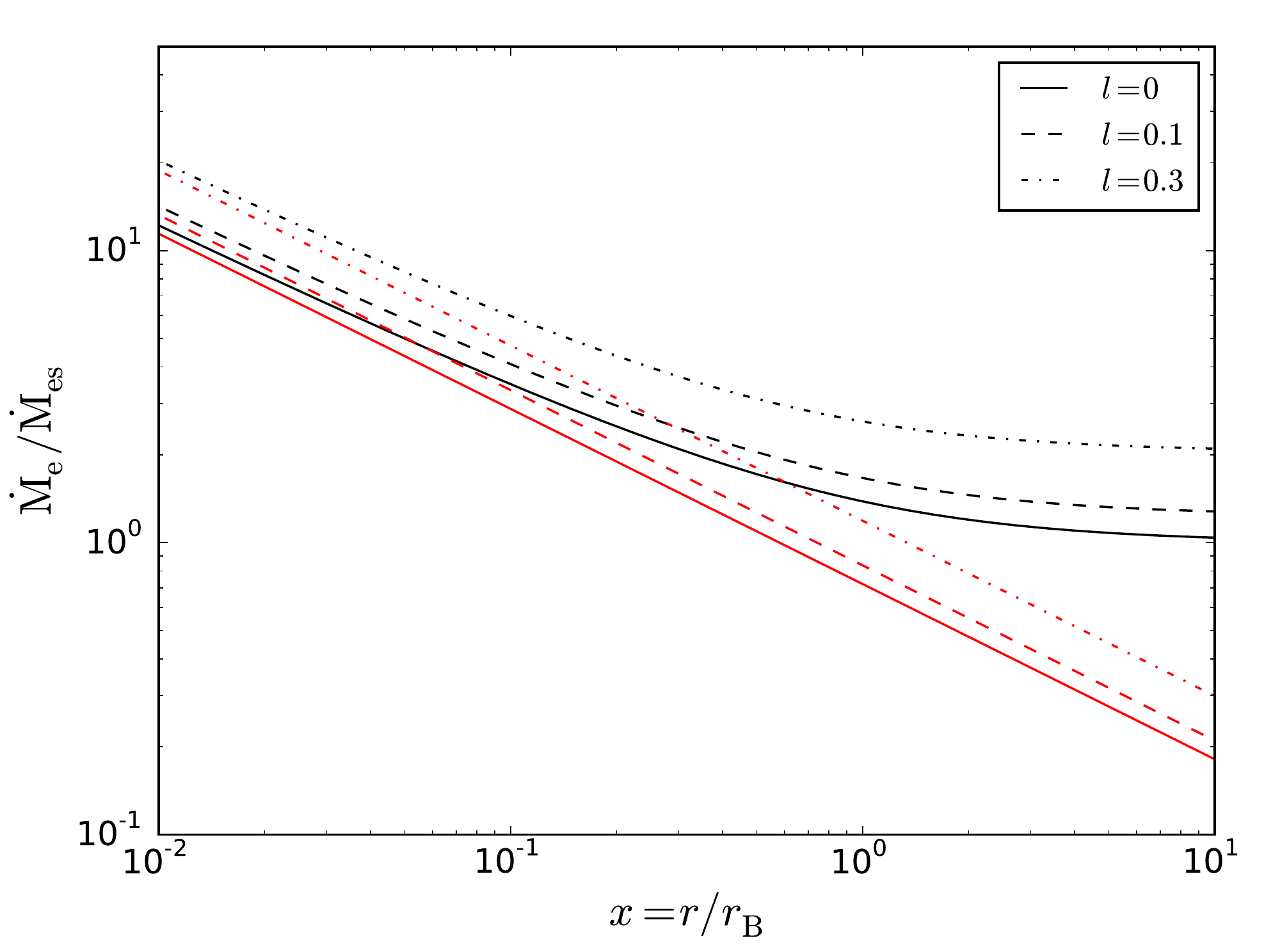}} \hfill 
        \caption{Bondi accretion model with electron scattering, for
          three values of $l=L/\Ledd$ indicated in the panels.  {\it
            Upper panel:} estimated  Bondi radius $\rbe$
          [eq. (33)], obtained from $T$ measured at a distance $r$
          from the MBH, as a function of $r$.  {\it Lower panel:} the
          estimated accretion rate $\Mdotbe$, in units of the true
          accretion rate, according to the Bondi theory, that includes
          electron scattering $\Mdotbes$ [eq. (34)], as a function of
          $r$. In both panels the black lines show the numerical
          solutions corresponding to different values of the
          normalized accretion luminosity, and the red ones the
          asymptotic solutions provided by eqs. (35)-(36);
          $\gamma=1.4$.}
 \label{fig:e-scatt}
\end{figure}

As well known, the Bondi solution is a purely hydrodynamical flow,
where heat exchanges are implicitly described by the polytropic index.
Therefore, for given polytropic index, and in absence of shock waves
(as for example in the transition from subsonic to
supersonic regime for the subcritical  $\lambda < \lambdacr$ case),
one could follow the entropy evolution of each fluid element along the
radial streamline, and determine the reversible heat exchanges.
However, in real accretion the energetics can be dominated by the
irreversible emission of energy near the MBH, that, for the radiative
component, is usually expressed as
\begin{equation} 
	L= \varepsilon\Mdotacc c^2,
\end{equation}
where $\varepsilon$ is the efficiency of the release of the accretion
energy, and $\Mdotacc$ is the mass accretion rate; in the
classical Bondi accretion, $\Mdotacc=\Mdotb$.  In principle
$\varepsilon$ can depend on luminosity $L$ or on $\Mdotacc$
(especially at low luminosities, as in the ADAF family [Narayan \& Yi
1995], and its variants).  At high accretion rates, the efficiency is
of the order of $\varepsilon_0=0.1$, and the accretion is likely
unsteady, so that Bondi accretion cannot be used (e.g., Ciotti \& Ostriker 2012 for
a review).  The emitted radiation interacts with the surrounding
medium and modifies the accretion process: the
radiation effects can be sufficiently strong to stop accretion (the
so-called negative feedback), and shut-off the central AGN.
Stationarity is almost impossible in these circumstances (Binney \&
Tabor 1995; Ciotti \& Ostriker 1997, 2001; Park et al. 2014).

However, when restricting to low accretion rates and considering only
electron scattering, in the optically thin
regime it is possible to generalize the classical Bondi accretion
solution by taking into account the radiation pressure effect (Taam,
Fu \& Fryxell 1991; Fukue 2001; Lusso \& Ciotti 2011).  In fact, for
electron scattering in the optically thin regime and in spherical
symmetry, the effective force experienced by a gas element can be
written as:
\begin{equation}
	F(r)= -{G \Mbh \rho(r)\chi\over r^2},\quad 
        \chi\equiv 1-l,\quad 
         l \equiv {L\over\Ledd},
\end{equation}
where $\Ledd=4\pi c G \Mbh\mpr /\sigma_{\rm T}$ is the Eddington
luminosity, and $\sigma_{\rm T}=6.65 \times 10^{-25}$cm$^2$ is the
Thomson cross section.  In the optically thin regime, being $l$
independent of radius, the radiation feedback can be implemented in
eq. (11) as a correction that reduces the gravitational force of the
MBH by the factor $\chi$; thus, for $\gamma>1$, the function $f$ in
eq. (11) becomes:
\begin{equation} 
	f(x)=x^{4 (\gamma-1)\over\gamma +1} 
               \left({\chi\over x} + {1\over \gamma-1}
      \right). 
\end{equation}
We do not give the analogous formula for the isothermal case (eq. 16),
because, as in the classical case, the relevant quantities can be
obtained by taking the limit for $\gamma\to 1^{+}$ of the formulae in
the general polytropic case.  One can repeat step by step the analysis
of Sect. 2, and show that the (unique) minimum of $f$ in eq. (26) is
reached for
\begin{equation} 
\xmin ={\chi (5-3\gamma)\over 4}.
\end{equation}
Being $\chi <1$, the position of the minimum moves inward with respect to
the classical case in eq. (13), and the value of $\fmin$ is just that
of the classical Bondi accretion (eq. 13) reduced by the factor
$\chi^{4(\gamma-1)/(\gamma+1)}$. Since the minimum of $g(\calM)$ is
independent of electron scattering, the critical value of the new
accretion parameter $\lambdaes$, at a given $\gamma$, is:
\begin{equation}
	\lambdaes = \chi^2\lambda,
\end{equation}
where $\lambda$ is the critical parameter in the corresponding classical
case.  Being $\chi <1$,  $\lambdaes$ is lower than in the
classical model. The true accretion rate, that
we now call $\Mdotbes$, is also reduced with respect to the classical
value $\Mdotb$, for given $\Mbh$, $\gamma$, and boundary conditions
at infinity:
\begin{equation} 
\Mdotbes = 4 \pi \rb^2 \lambdaes\rhoinf\csinf = \chi^2 \Mdotb. 
\end{equation}
Note that $\Mdotbes$ enters the value of $\chi$ through the
accretion luminosity $L$, so that eq. (29) can be seen as an implicit
equation for $\Mdotbes$.  Remarkably, this equation can be explicitly
solved by introducing the Eddington mass accretion rate
\begin{equation}
\Mdotedd\equiv{\Ledd\over \varepsilon_0c^2},\quad l = {\Mdotbes\over\Mdotedd} {\varepsilon\over\varepsilon_0}. 
\end{equation}
In general $\varepsilon$ depends on the accretion rate, but in the
following for illustrative purposes we assume
$\varepsilon=\varepsilon_0$. From eqs. (29)-(30), one
obtains explicitly $\Mdotbes/\Mdotedd$ in terms of
$\Mdotb/\Mdotedd$, by solving the quadratic equation:
\begin{equation} 
\Mdotbes= \left(1- {\Mdotbes\over\Mdotedd} \right)^2\,\Mdotb 
\end{equation}
(see Lusso \& Ciotti 2011). For example, at low accretion rates
$\Mdotbes \sim \Mdotb$, while when $\Mdotb$ would diverge
to infinity, one reaches the asymptotic accretion rate
\begin{equation}
{\Mdotbes\over\Mdotedd} \sim 1 - \sqrt{\Mdotedd\over\Mdotb}. 
\end{equation}

We now apply to the Bondi solution with electron scattering the same
procedure of Sect. 2.1, to quantify the differences, as a function of
radius, between the true ($\rb$) and estimated ($\rbe$) Bondi radius,
and the true ($\Mdotbes$) and estimated ($\Mdotbe$) accretion rate,
where $\rbe$ and $\Mdotbe$ are defined as in eq. (19).  It is easy to
show that:
\begin{equation} 
{\rbe(x)\over\rb}= \tilde\cs(x)^{-2}
=\rhotil(x)^{1-\gamma}=\left({x^2 \calM\over \lambdaes}\right)^{2(\gamma-1)\over\gamma+1},
\end{equation}
\begin{equation}
{\Mdotbe(x)\over\Mdotbes} = 
{\lambda\over\lambdaes}\left[{\rb\over\rbe(x)}\right]^{5-3\gamma\over
  2(\gamma-1)}= 
{1\over\chi^2} \left[{\rb\over\rbe(x)}\right]^{5-3\gamma\over 2(\gamma-1)},
\end{equation}
where the density and temperature profiles are now those appropriate
for accretion with electron scattering.  For $x \to \infty$, by
definition, $\rbe\to\rb$. Again, as for the classical Bondi problem in
Sect. 2.1, $\rbe$ is always smaller than $\rb$ for $\gamma >1$, since
the gas sound speed increases inward\footnote{This trend for the
  radial behavior of $T$ is easily understood when considering that
  the problem with electron scattering is just the classical Bondi
  problem on a MBH of reduced mass equal to $\chi\Mbh$.}; for $\gamma =1$, $\rbe=\rb$ 
 independently of the distance $r$ from the center.

For the mass accretion rate, again in analogy with what found for the
classical Bondi problem, we have that
$\Mdotbe(x)\to\Mdotbes/\chi^2=\Mdotb$ for $x\to\infty$; that, for
$\gamma=5/3$, $\Mdotbe(x)=\Mdotbes/\chi^2$ independent of $r$; and
that, for $\gamma =1$, $\Mdotbe (x)
=\rhotil(x)\Mdotbes/\chi^2=\rhotil(x)\Mdotb$.  Note that now the resulting bias depends
not only on the distance $r$ where the density and temperature are taken,
but also on the value of $\Mdotbes/\Mdotedd$, through the
parameter $\chi$.

The asymptotic analysis shows that near the center:
\begin{equation}
{\rbe(x)\over \rb}\sim \chi^{3(1-\gamma)\over 2} 
\left({\sqrt{2}\over\lambda}\right)^{\gamma-1}
x^{3(\gamma-1)\over 2},\quad  x \to 0^+,
\end{equation}
\begin{equation}
{\Mdotbe(x)\over\Mdotbes}\sim \chi^{7-9\gamma\over 4}
\left({\lambda\over\sqrt{2}}\right)^{5-3\gamma\over 2} 
x^{-{3(5-3\gamma)\over 4}},\quad  x \to 0^+. 
\end{equation}
In particular, the r.h.s. in eqs. (35)-(36) are just the r.h.s.  of
eqs. (22)-(23) for the classical Bondi problem, multiplied by
$\chi^{3(1-\gamma)\over 2}$ and $\chi^{7-9\gamma\over 4}$,
respectively. Therefore, $\rbe/\rb$ again decreases for $\gamma >1$
and $x\to 0$. For $\gamma =5/3$, $\rbe$ scales linearly with $r$,
as $\rbe\sim 2^{5/3}r/\chi$, and again $\rbe >r$, but by  a
larger factor than in classical Bondi accretion.  For 
$\gamma <5/3$, $\Mdotbe$ overestimates the true accretion rate
$\Mdotbes$; for $\gamma =1$, $\Mdotbe\propto x^{-3/2}$.  Again,
as in the classical Bondi problem, it is possible to recover the true
accretion rate from $\rbe$ and $\Mdotbe$  near the center, as
$\Mdotbes\sim \sqrt{\chi}(\sqrt{2}/\lambda) (r/\rbe)^{3/2}\Mdotbe
(r)$. The resulting quadratic equation for $\Mdotbes$ can be easily
solved after writing the correction coefficient $\chi$ in
terms of $\Mdotbes$, following the procedure described below eq. (30).

The properties above are illustrated by Fig.~2, that shows the
numerical and asymptotic solutions for $\rbe/\rb$ [eqs. (33) and
(35)], and for $\Mdotbe/\Mdotbes$ [eqs. (34) and (36)], as a function
of the distance $r$, for different values of $l$, and $\gamma$ fixed
to the representative value 1.4.  The figure gives a quantification of
the bias on the estimates of the Bondi radius and mass accretion rate:
$\rbe$ is always an underestimate of $\rb$, as expected, while
${\Mdotbe}$ is always an overestimate of the true accretion rate, even
by a large factor if $r<0.1\rb$ (and, of course, increasing for larger
$l$).

\section{Bondi accretion with electron scattering onto MBHs at the
  center of galaxies} 
\label{sec:Bondi+gal}

We can now discuss the full problem, i.e., we investigate how standard Bondi accretion
is modified by the additional potential of the host galaxy, and by
electron scattering. We then generalize the
previous conclusions about the fiducial Bondi radius $\rbe$ and 
mass accretion rate $\Mdotbe$ obtained using quantities at finite distance
from the MBH.

We assume spherical symmetry for the host galaxy, so that its gravitational
potential can be written in full generality as
\begin{equation}
\phig = - {G\Mg\over\rg}\,\psi\left({r\over\rg}\right),
\end{equation}
where $\Mg$, $\rg$ and $\psi$ are respectively the total galaxy mass,
a characteristic scale-length, and the
dimensionless galaxy potential.  By introducing the parameters
\begin{equation} 
\calR \equiv {\Mg\over\Mbh},\quad
\xi \equiv {\rg\over\rb},
\end{equation}
the effective total gravitational potential to be inserted in the expression for the function $f$ can be written as:
\begin{equation} 
\phit = - {G\Mbh\over\rb} \left[{\chi\over x} + {\calR\over\xi}\, \psi\left({x\over\xi} \right)\right].
\end{equation} 
The addition of the galaxy potential $\phig$ changes the function $f$
in eqs. (11) and (16), and consequently also the values of
$\fmin(\chi,\calR,\xi)$ and of the critical $\lambda$ (that now we
call $\lambdat$); the function $g(\calM)$ is unaffected by the
addition of the galaxy.  Thus, $\lambdat$ is now given in the
polytropic case by eq. (12) with the new value of $\fmin$, and again
the isothermal case can be obtained as a limit for $\gamma\to 1^+$ of
the polytropic problem.  For a generic galaxy model, it is no longer
possible to obtain an analytical expression for $\xmin$, $\fmin$, and
$\lambdat$, and they must be determined numerically.  As we will see
in Sect. 5, the galaxy potential can produce more than one minimum for
the function $f$; in this case it is easy to conclude that the general
considerations after eqs. (11) and (16) refer to the {\it absolute}
minimum of $f$, and so $\xmin$ gives - along the critical solution -
the location of the {\it sonic point}\footnote{It can be shown (e.g.,
  Frank, King, \& Raine 1992) that  the sonic radius $r_s$ obeys the
  identity $\cs^2(r)=(r/2)\,d\phi_{\rm t}/dr$ where $\phi_t$ is the
  total gravitational potential.}. Of course, when $\calR\to 0$ (or
$\xi\to\infty$), the galaxy contribution vanishes,
$\lambdat=\lambdaes$, and the formulae of Sect. 3 hold.  Instead, by
setting $\chi=1$, one can determine the sole gravitational effects of
the host galaxy. In Appendix A we describe how to compute the
correction terms for $\lambda$ due to the presence of a generic galaxy
model, in some special cases, while in Appendix B an important
property of monoatomic adiabatic accretion in generic galaxy
potentials is derived, i.e., that $\lambdat =\lambdaes=\chi^2/4$,
where the last identity comes from eq. (28) with $\gamma=5/3$.

\subsection{Mass accretion estimates}

Following the procedure described in Sects. 2 and 3, we evaluate 
the deviation of $\rbe$ and $\Mdotbe$  from the true values
of $\rb$ and of the mass accretion rate, that we now indicate as $\Mdott$.  The
presence of a galaxy changes the accretion rate on the central MBH;
eq. (9) still holds, but now:
\begin{equation} 
	\Mdott = 4 \pi \rb^2 \lambdat \rhoinf
       \csinf={\lambdat\over\lambda}\Mdotb,
\end{equation}
where $\Mdotb$ is the Bondi accretion rate, for the same chosen
boundary conditions $\rhoinf$ and $\csinf$, in absence of the galaxy
and of radiation pressure. The expressions for $\rbe$ and $\Mdotbe$
are given by the analogous of eqs. (20)-(21) and (33)-(34), where now
the physical variables are taken along the solution in presence of the
galaxy and of electron scattering. From eq. (19) then one has:
\begin{equation} 
{\rbe(x)\over\rb}= \tilde\cs(x)^{-2}
=\rhotil(x)^{1-\gamma}=\left({x^2 \calM\over \lambdat}\right)^{2(\gamma-1)\over\gamma+1},
\end{equation}
and
\begin{equation} 
{\Mdotbe(x)\over\Mdott}=  {\lambda\over\lambdat}
\left[{\rb\over\rbe (x)}\right]^{5-3\gamma\over 2(\gamma-1)}.
\end{equation}
For $x \to \infty$, by definition, $\rbe\to\rb$, and,
$\Mdotbe(x)\to\Mdott \lambda/\lambdat=\Mdotb$.  As in the previous
simpler accretion cases, for $\gamma =1$ one has that $\rbe=\rb$,
independently of the distance from the center, and
$\Mdotbe(x)=\rhotil(x)\Mdott\lambda/\lambdat=\rhotil(x) \Mdotb$.  For
$\gamma=5/3$, and from the result of Appendix B,
$\Mdotbe(x)=\Mdott/\chi^2$, independent of $r$. Additional properties
of $\lambdat$ will be discussed in the next Section.

\begin{figure*}
        \centering
	 \includegraphics[height=0.5\textwidth, width=1\textwidth]{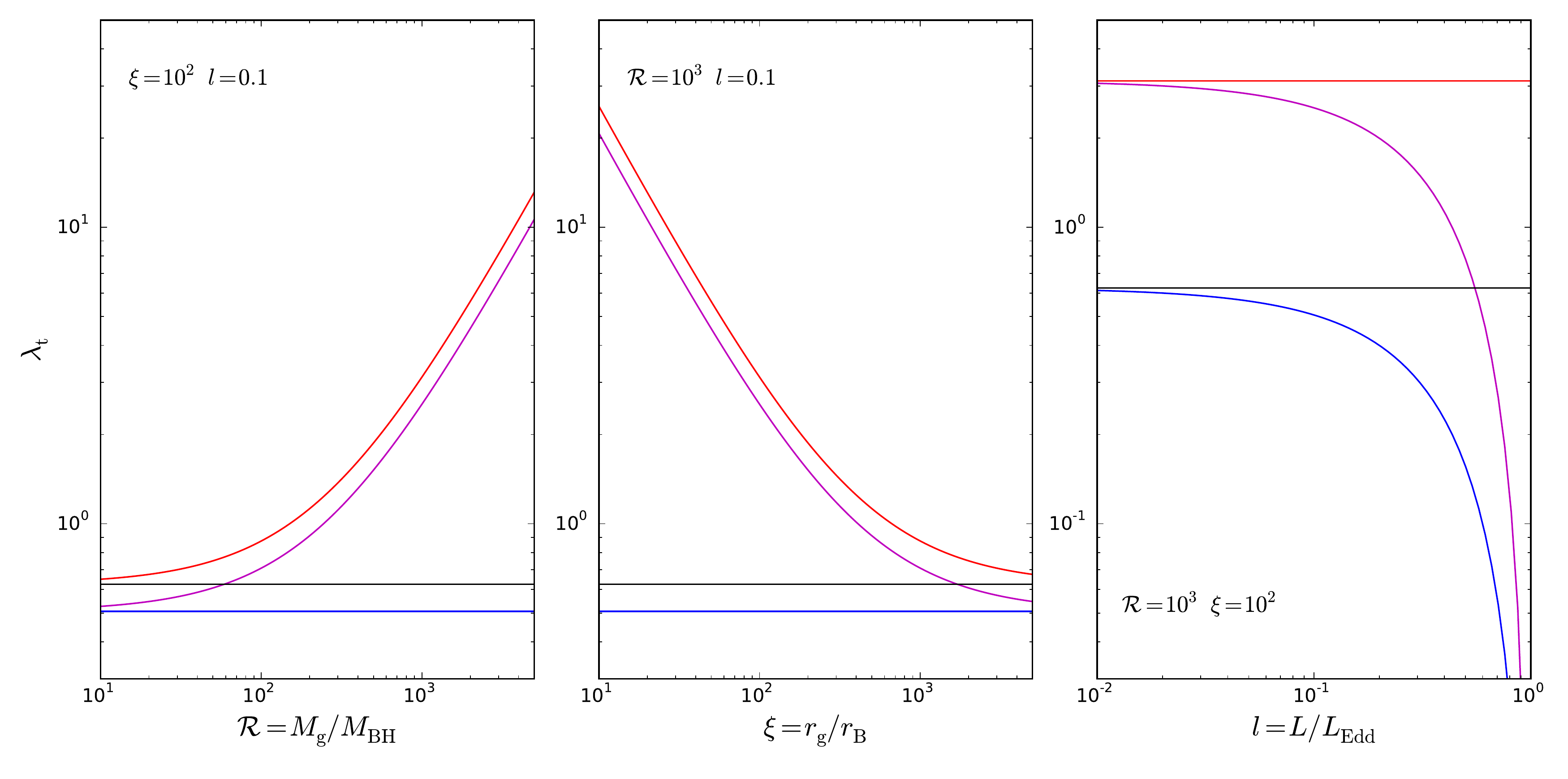}
         \caption{Bondi accretion model with electron scattering at
           the center of a Hernquist galaxy: the magenta line shows
           $\lambdat$ as a function of $\calR=\Mg/\Mbh$ (left panel),
           of $\xi=\rg/\rb$ (middle panel), and of $l$ (right
           panel). $\lambdat$ is also compared to $\lambda$ of the
           classical Bondi model (in black), to that of the Bondi
           model with a Hernquist galaxy (in red), and to $\lambdaes$
           of the Bondi model with electron scattering (in blue).  The
           polytropic index is $\gamma=1.4$.  As expected, $\lambdaes
           < \lambda$, because the electron scattering lowers
           $\lambda$ [see eq. (28)]; similarly, $\lambdat$ is always
           lower than the red line corresponding to the galaxy only
           [see also eqs. (49)-(50)]. }
       \label{fig:Ltot}
\end{figure*}

Repeating the asymptotic analysis of Sects. 2 and 3, near the center
we now derive:
\begin{equation}
{\rbe(x)\over\rb} \sim \chi^{\gamma -1\over 2}
\left({\sqrt{2}\over\lambdat}\right)^{\gamma-1}
x^{3(\gamma-1)\over 2},\quad  x \to 0^+,
\end{equation}
\begin{equation}
{\Mdotbe(x)\over\Mdott} \sim \chi^{-{5-3\gamma\over 4}}
{\lambda\over\sqrt{2}}
\left({\sqrt{2}\over\lambdat}\right)^{3(\gamma-1)\over 2} 
x^{-{3(5-3\gamma)\over 4}},\quad x \to 0^+.
\end{equation} 
As in the other cases, $\rbe/\rb$ decreases for $\gamma>1$ and $x\to
0$. For $\gamma=5/3$, $\rbe$ scales linearly with $r$, as
$\rbe\sim2^{5/3} r/\chi$ (because, for $\gamma=5/3$,
$\lambdat=\lambdaes=\chi^2/4$, see Appendix B), and again $\rbe>r$ for
$r\to 0$, but by a larger factor than in classical Bondi accretion.
Remarkably, near the center it is again possible to obtain the true
$\Mdott$ by using the fiducial $\Mdotbe$ and $\rbe$. In fact, for
$x\to 0$, $\lambdat$ cancels out from eqs. (42)-(43), and $\Mdott\sim
\sqrt{\chi}(\sqrt{2}/\lambda) (r/\rbe)^{3/2}\Mdotbe (r)$. This
quadratic equation can be easily solved for $\Mdott$ as in Sect. 3,
after expressing the radiative correction coefficient $\chi$ in term
of $\Mdott$, following the procedure described below eq. (30).

Before proceeding with the solution of this accretion problem, we need
to examine whether $\Ledd$ changes in presence of a galaxy. If this is
the case, then the dimensionless accretion luminosity $l=L/\Ledd$ and
the associated $\chi$ would be a function of the specific model
investigated.  However, it can be shown that for any given galaxy
model, characterized by the cumulative mass distribution $\Mg(r)$, the
Eddington luminosity depends on radius as
\begin{equation} 
	\Ledd(r) =  \left[ 1 +{\Mg(r)\over\Mbh} \right]\Ledd,
\end{equation}
where $\Ledd$ is the classical value whose expression is given below
eq. (25). Therefore, $\Ledd (r)$ is minimum at the center, with
$\Ledd(0)=\Ledd$, and steady accretion - also in presence of a galaxy
- cannot exist for $L>\Ledd$. This implies that $l$ and $\chi$, also
in presence of a galaxy, are still defined as in Sect. 4.

\begin{figure}
       \subfloat{\includegraphics[height=0.45\textwidth, width=0.48\textwidth]{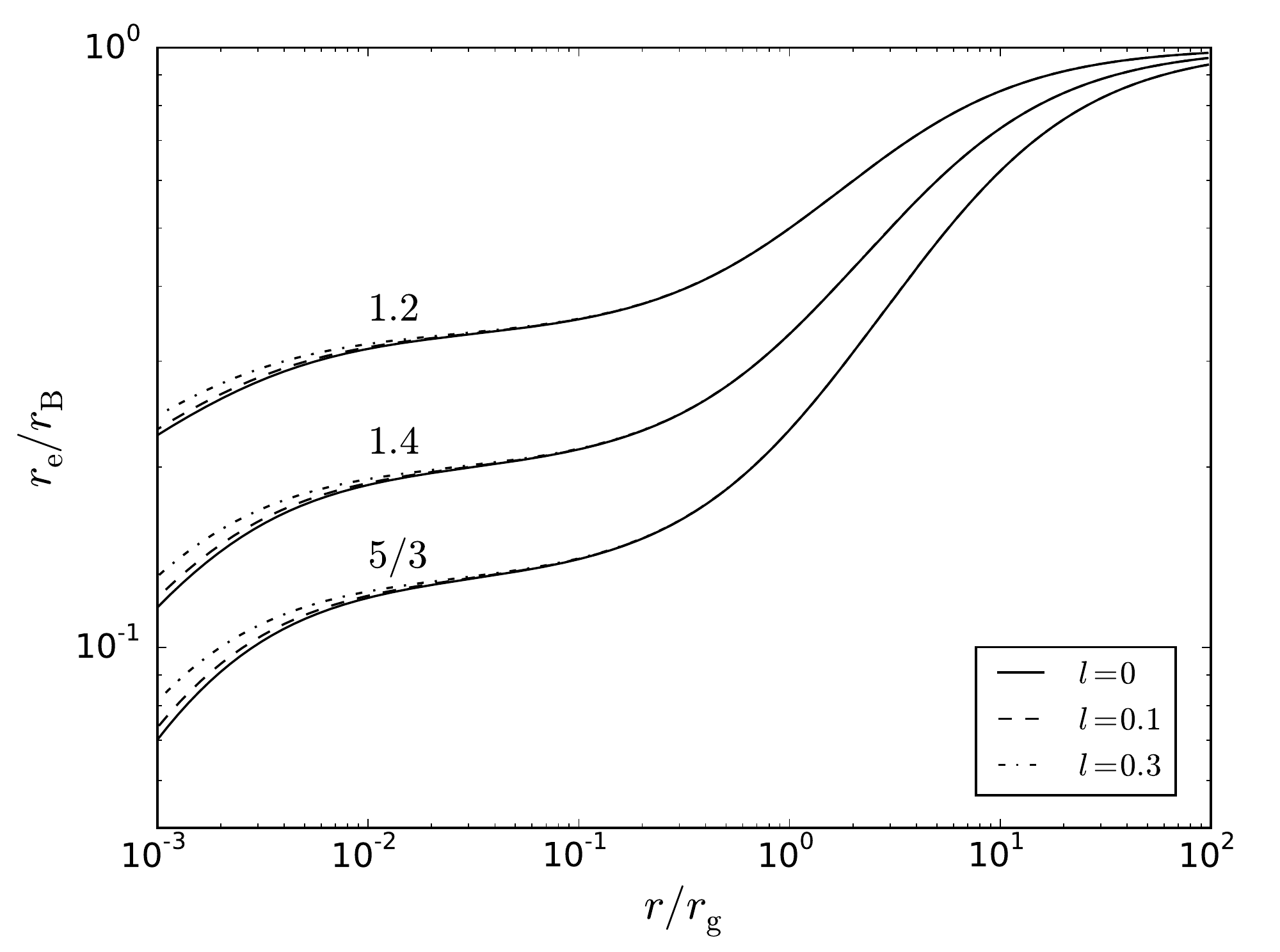}} \\
	\subfloat{\includegraphics[height=0.45\textwidth, width=0.48\textwidth]{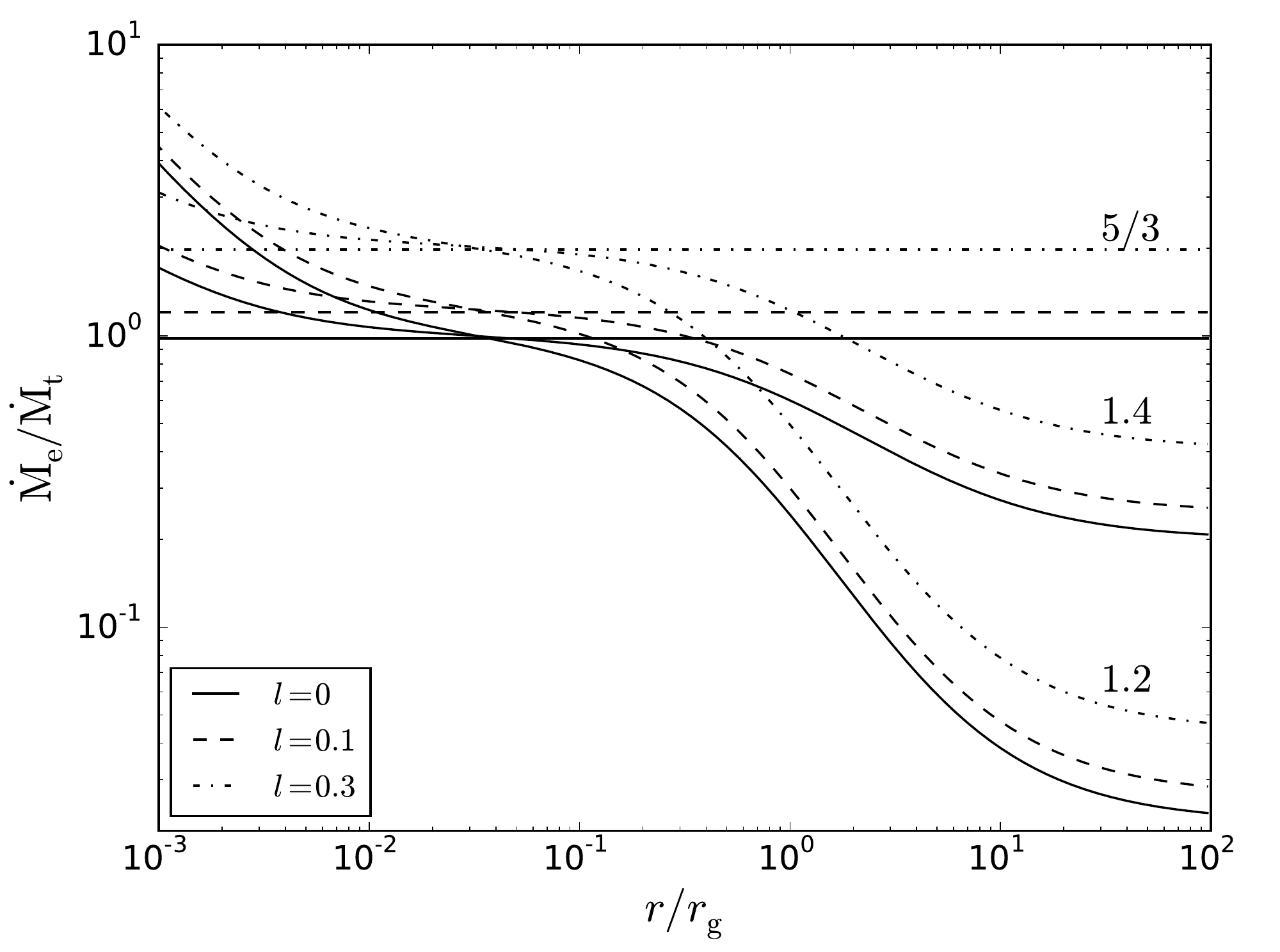}} 
       \caption{Bondi accretion model with electron scattering at the
          center of a Hernquist galaxy. {\it Upper panel:} estimated
          value of the Bondi radius $\rbe$ [eq. (41)], obtained from
          $T$ measured at a distance $r$ from the MBH.  {\it Lower
            panel:} estimated accretion rate $\Mdotbe$, in units of
          the true accretion rate $\Mdott$ [eq. (42)], as a function
          of $r$. In both panels $l=0$ (solid line), or $l=0.1$
          (dashed line), or $l=0.3$ (dotted line); $\calR=10^3$ and
          $\xi=100$.  Note that, for each $l$, a different $\Mdott$ is
          implied, according to eq. (40) [see also eq. (50) for an
          explicit dependence of $\lambdat$ on $l$, in a limiting
          case].  }
      \label{fig:biastot}
\end{figure}

\section{The case of a Hernquist galaxy model} 
\label{sec:Hernquistgal}

In order to provide quantitative estimates for the trends of
$\rbe(x)/\rb$ and $\Mdotbe(x)/\Mdott$, we need to adopt a specific
galaxy model, to determine $\lambdat$.  As already noticed, the value
of $\lambdat$ is known once the absolute minimum $\fmin(\chi,\calR,
\xi)$ is known; this, in turn, requires the determination of $\xmin$.
As a  realistic galactic potential we consider that corresponding to the
Hernquist (1990) density profile, that describes well the mass distribution of
early-type galaxies; however, we stress that several results below
remain  true also for other galaxy models, such as the
so-called $\gamma$-models (Dehnen 1993, Tremaine et al. 1994).  The
gravitational potential of the Hernquist model is:
\begin{equation} 
	\phig=-{G\Mg\over r+\rg},
\end{equation}
where the scale-length $\rg$ is related to the galaxy effective radius
$R_{\rm e}$ as $\rg \simeq R_{\rm e}/1.82$. Thus, from eq. (39) for
$\gamma>1$ one has:
\begin{equation} 
	f(x)=x^{4 (\gamma-1)\over \gamma +1}  \left({\chi\over x} +
          {\calR\over x+\xi}  + {1\over \gamma-1} \right),
\end{equation}
and the formula for the isothermal case is obtained from eq. (16).

\subsection{Analytical results }

Before presenting the numerical results for the Hernquist galaxy
model, a preliminary discussion is useful in order to obtain hints on
the expected behavior of the solution. For the Hernquist model the
analysis is sufficiently simple to prove some interesting results, yet
the case already illustrates the difficulties encountered in accretion
problems in galaxy potentials. Basically, the following analysis
focuses on the behavior (number and location) of the critical points
of $f$, on the determination of the absolute minimum $\fmin$, and
finally on the dependence of $\lambdat$ on the model parameters.

The full analysis of the number and location of minima of eq. (47) is
given in Appendix C, where eq. (C1) gives the expression for $f'$ when
$\gamma >1$; for $\gamma =1$, $f'$ is obtained as the limit for
$\gamma\to 1^+$ of the expression for $f'$ in eq. (C1).

From the expression for $f(x)$ one sees that in three cases the
determination of $\xmin$ and $\fmin$ is trivial. The first is when
$\xi \to \infty$ (or $\calR \to 0$), then the galaxy contribution
vanishes, and eqs.~(27)-(28) are recovered.  The second case
corresponds to $\xi=0$, when the problem reduces to Bondi accretion
onto a MBH of mass $(\chi+\calR)\Mbh$. For these two cases the
position of the only minimum of the function $f$ (i.e., of the sonic
radius), and the critical value $\lambdat$, for $1 \le \gamma \le
5/3$, are given by:
\begin{equation} 
\xmin = {(\chi+\calR) (5-3\gamma)\over 4},\quad \lambdat =
(\chi+\calR)^2 \lambda,
\end{equation}
where $\lambda$ is the critical value for the corresponding classical
Bondi problem (Sect. 2).  A third simple case is when $\xi\geq 0$ but
$\gamma=5/3$: $f'$ is definite positive [eq. (C3)], so that the
minimum of $f$ is reached at the center (as for the adiabatic Bondi
problem without the galaxy).  Simple algebra shows that the Hernquist
model with $\xi>0$ obeys the (very weak) hypoteses of Appendix B, and
so the general identity $\lambdat=\chi^2/4$ holds. Note that this
value differs from that in eq.~(48) pertinent to the $\xi=0$ case: in
fact, in Appendix B we also show that the $\gamma=5/3$ case for the
Hernquist model is singular for $\xi\to 0$, as also manifested by the
fact that $\lim\limits_{\xi \to 0} \fmin = \chi$ while
$\fmin(\xi=0)=\chi+\calR$. As a strictly related result, again from
Appendix B, it follows that accretion in the $\gamma=5/3$ case is
impossibile in a generic galaxy models without a central MBH, because
$\fmin=0$ and so $\lambdat=0$ (as confirmed by setting $\chi=0$ in the
identity above).

As shown in Appendix C, for accretion in the general Hernquist galaxy
model with central MBH and electron scattering, $\xmin$ can be
obtained by solving an algebraic (cubic) equation, thus providing in
principle the explicit expression for $\fmin$ and $\lambdat$, at
variance with the common situation encountered for other galaxy
models. The solution presents no difficulites, however it is quite
cumbersome, and here we just recall that, depending on the specific
values of $\calR$, $\xi$ and $\gamma$, there can be a single minimum
for $f$, or two minima and one maximum.  In Appendix C we provide
analytical formulae that can be easily implemented in numerical
studies, allowing to determine, for any given choice of $\calR$, $\xi$
and $\gamma$, the number and the location of the critical points of
$f$.

Altough the explicit expression for $\lambdat$ can be easily
constructed, here we just give some hints on its behavior by
considering cases of perturbative analysis (namely, for $\xi\to\infty$
or $\calR\to 0$, and for $\xi\to 0$), along the lines described in
Appendix A (eq. A6 and following for details). For example, for
$\gamma=1$, the results in Appendix C3 show that there is only one
critical point for $f$ in the considered cases, and expanding around
the unperturbed minimum (whose position is given in eq. 27 for
$\xi\to\infty$ or $\calR\to 0$, and in eq. 48 for $\xi\to 0$), one
finds that the leading term of $\lambdat$ is:
\begin{equation} 
\lambdat\sim\lambda \times 
\begin{cases}
\displaystyle{\chi^2 e^{\calR\over \xi+\chi/2},\quad \xi\to\infty, \quad\calR\to 0,}
\\ \\ 
\displaystyle{(\chi+\calR)^2e^{-{4\calR\xi\over (\chi+\calR)^2}},\quad \xi\to 0,}
\end{cases}
\end{equation}
where $\lambda=e^{3/2}/4$.

A similar analysis can be performed for $1<\gamma < 5/3$.  A careful
expansion around the absolute minimum of $f$, as determined from
Appendix C, shows that for $\xi\to\infty$ (or $\calR\to 0$), and for
$\xi\to 0$ we have respectively:
\begin{equation}
\lambdat\sim\lambda\times 
\begin{cases}
\displaystyle{
\chi^2 \left[1 +{4(\gamma-1)(5-3\gamma)\calR\over 
    (\gamma+1)[\chi (5-3\gamma)+4\xi]}\right]^{\gamma+1\over 2(\gamma-1)},}
\\ \\
\displaystyle{(\chi +\calR)^2 \left[1-{16\ \calR\ \xi (\gamma-1)\over
      (1+\calR)^2 (\gamma+1) (5-3\gamma)} \right]^{\gamma+1\over 2(\gamma-1)}. }
\end{cases}
\end{equation}
Note how eq.~(50) for $\gamma \to 1$ recovers both cases of
eq. (49). As expected, however, the second of eq.~(50) is singular for
$\gamma\to 5/3^-$, for the reasons described in Appendix B.  Note
that, at variance with the Bondi accretion with electron scattering
only, now the electron scattering coefficient $\chi$ cannot be
factorized in the expression for $\lambdat$, i.e., $\lambdat$ in
general cannot be factorized as eq.~(28) as the product of the
electron scattering coefficients times a function relative to the
galaxy accretion without electron scattering effects. It follows that
in presence of a galaxy the procedure described in Sect. 4
[eqs. (31)-(32)] cannot be applied analytically.

We stress that all the formulae above have been also derived by direct
expansion of the exact solution of eq.~(C2), thus providing an
independent test of the expansion procedure in Appendix A.

\subsection{Numerical results}

For a Hernquist galaxy potential it is not possible to obtain the
analytical expression for the critical accretion parameter $\lambdat$;
thus, we determined numerically the density and temperature profiles
of the Bondi model with electron scattering and a Hernquist galaxy,
and calculated $\lambdat$, $\rbe$ and ${\Mdotbe}/{\Mdott}$. The
results for $\lambdat$ are plotted in Figure \ref{fig:Ltot}, where we
show the trend of $\lambdat$ with respect to $\calR=\Mg/\Mbh$, for
fixed $\xi=100$ and $l=0.1$, with respect to $\xi=\rg/\rb$ for
$\calR=10^3$ and $l=0.1$, and finally with respect to $l$ for
$\calR=10^3$ and $\xi=100$.  In the left panel, for a fixed $\xi$, for
which we assumed a realistic value\footnote{When $\rb$ is of the order
  of few tens of pc, as expected for the MBHs at the center of early
  type galaxies, the value $\xi=100$ corresponds to $\rg$ of few kpc,
  i.e., a reasonable value for $\rg\simeq R_e/1.82$.} of 100,
$\lambdat$ increases as $\calR$ increases, i.e., as the galaxy total
mass increases.  In the middle panel, by setting $\calR=10^3$, as
dictated by the Magorrian relation (Magorrian et al. 1998), we see
that $\lambdat$ decreases as $\xi$ increases, since the effect fo the
galaxy is vanishing (and in fact $\lambdat$ reaches down to
$\lambdaes$). Finally, in the right panel $\lambdat$ goes to zero for
increasing values of $l$. Note that for the adopted, illustrative
values of $\calR$ and $\xi$, the analysis in Appendix C3 (see also
Fig. C1) shows that there are three critical points for $f$ (two
minima and one maximum), so that $\lambdat$ is obtained from the
absolute minimum.

These results for $\lambdat$ are also compared to those obtained for
the previous simpler models, i.e., the $\lambda$ of the classical
Bondi model, the lower $\lambdaes$ of the Bondi model with electron
scattering, and that of the Bondi model with only the Hernquist
galaxy. The latter keeps always larger than $\lambdat$, since the
effect of the electron scattering is to lower $\lambda$ (eq. 28); of
course, it tends to $\lambda$ of the classical Bondi model for $\calR
\to 0$ (i.e., for $\Mg \to 0$), and for $\xi \to \infty$ (i.e., very
large $r_{\rm g}$, and then negligible effect of the galaxy).  Also,
$\lambdat \to \lambdaes$ for $\calR \to 0$ and for $\xi \to \infty$
(both cases in which the effect of the galaxy potential is vanishing).

In Fig.~4 we represent $\rbe$ and ${\Mdotbe}/{\Mdott}$, as a function
of $r/\rg$, for different values of $\gamma$ and $l$.  The figure
presents a quantification of the bias on the estimates of the Bondi
radius and of the true mass accretion rate. Even in this more complex
case, $\rbe$ provides an underestimate of $r_{\rm B}$.  For
$\gamma=5/3$, ${\Mdotbe}={\Mdott}$ if $l=0$, otherwise ${\Mdotbe}$ is
always an overestimate of the true accretion rate (increasing for
larger $l$). For the other $\gamma$'s, ${\Mdotbe}$ provides an
overestimate for $r<\rg$ (more severe for larger $l$), and an
underestimate for $r>\rg$ (less severe for larger $l$).

\section{Summary and conclusions}

Due to its simplicity, the classical Bondi accretion theory remains
the standard paradigm against which more realistic descriptions are
confronted, and observations are interpreted. Also, Bondi accretion is
adopted to get a simple approximation for the mass accretion rate in
semi-analytical models and numerical simulations, particularly when
numerical resolution is not high enough to probe in a self-consistent
way the regions near the central MBH.  Given the wide use of the
classical Bondi theory, and considered that accretion on the MBHs at
the center of galaxies is certainly more complicated than the
description provided by this theory, the motivation for this work was
to generalize the theory, including the radiation pressure feedback
due to electron scattering in the optically thin approximation, and
the effect of a general gravitational potential due to a host
galaxy. All the hypotheses of classical Bondi accretion (stationarity,
absence of rotation, spherical symmetry) were maintained.  In
addition, the present work provides a quantititative answer to a major
question, namely what is the bias induced on the estimates of the
Bondi radius and mass accretion rate when adopting as bona-fide values
of the hydrodynamical variables their values at some finite distance
from the center. This issue is relevant for observational (e.g.,
Pellegrini 2005, 2010; Russell et al. 2015) and numerical works (e.g.,
Di Matteo et al. 2005, Park et al. 2014, DeGraf et al. 2015).  The
main results can be summarized as follows.

1) For the three models of classical Bondi accretion, of accretion
with electron scattering, and of accretion on a MBH at the center of a
galaxy with electron scattering, we provide the exact formulae for
$\rbe/\rb$ (where $\rb$ is fixed for a chosen value of $T_{\infty}$),
as a function of the distance $r$, the Mach number ${\cal M}(r)$, and
the accretion parameter; these can be used when the temperature is
taken at any $r$ along the ``modified'' Bondi solution. We then give
the exact formulae for the mass accretion bias (respectively
$\Mdotbe/\Mdotb$, $\Mdotbe/\Mdotbes$, and $\Mdotbe/\Mdott$)
in terms of the bias on the Bondi radius ($\rbe/\rb$). For a
quantitative estimate of the bias, the knowledge of the numerical
solution of the associated accretion problem is required.

2) The formulae above allow to get some general results without
previous knowledge of the exact numerical solution, for particular
$\gamma$ values. In the monoatomic adiabatic case ($\gamma=5/3$),
$\Mdotbe$ is a constant fraction of the true mass accretion rate
($\Mdotb$, or $\Mdotbes$, or ${\rm \dot{M}_t}$), independently of the
distance $r$ from the MBH.  The fraction is exactly unity in case of
classical accretion ($\Mdotbe=\Mdotb$), and it is given by the ratio
of the critical accretion parameters in the other cases: with
radiation pressure, $\Mdotbe=\Mdotbes\lambda/\lambdaes=
\Mdotbes/\chi^2$; with electron scattering and a galaxy potential,
$\Mdotbe=\Mdott\lambda/\lambdat=\Mdott/\chi^2$ (see
Sect. 5.1 for more details and the particular case of a Hernquist
galaxy).  In the isothermal ($\gamma=1$) case, there is no bias in the
estimated value of the Bondi radius ($\rbe=\rb$), independently of the
distance from the center; the bias provided by $\Mdotbe$ is
proportional to the density of the accreting material.

3) The trends of $\Mdotbe$ and $\rbe$ near the center come from the
asymptotic expansion of the solutions for $\calM (r)$ and $f(r)$: the
bias can be written in terms of the distance $r$ and of the value of
the critical accretion parameter. In particular, for $\gamma >1$,
$\rbe$ provides a larger and larger underestimate of $\rb$ as $r$ is
approaching the center [in all cases $\rbe/\rb\propto
(r/\rb)^{3(\gamma-1)/2}$].  $\Mdotbe$ correspondingly provides an
overestimate of $\Mdotb$, $\Mdotbes$, and $\Mdott$.

4) From the asymptotic expansion of the solutions near the center, for
all three accretion models considered, it is also shown how to recover
the true value of the mass accretion rate from $\rbe$ and $\Mdotbe$
measured at some distance $r$ from the center. These formulae are
useful for observational and numerical works.

5) When the analytical solution is known for an accretion case, a
general asymptotic technique is given that allows to obtain the
correction terms for the value of the critical accretion parameter for
slightly different models, avoiding the need of a numerical solution
of the accretion problem (Appendix A).

6) In the illustrative case of a Hernquist galaxy model, the
determination of $\xmin$ and $\lambdat$ reduces to the study of a
cubic equation.  The case is fully discussed analytically (Appendix
C), and it is shown that more than one critical point for $f$ can
exist. We provide a simple analytical framework to determine whether
one or two minima occur, and the correction terms for $\lambdat$,
determined with the method of point 5).

7) We finally solved numerically the accretion problem for the three
models considered, to provide quantitative estimates of the bias.  The
analytical formulae turned out to be in perfect agreement with the
numerical results.

This work reveals how using values of density and temperature at
various radii when deducing accretion properties of MBHs at the center
of galaxies produces values of the mass accretion rate that should be
taken with care. We here obtained, however, results sufficiently
general to provide correction factors to be used in various
observational and numerical works.

\appendix

\section{Asymptotic expansion for the accretion parameter} 
\label{app:A}

>From the discussion in Sect. \ref{sec:class} it follows that the
critical value of the accretion parameter $\lambda$ in the Bondi
theory is given by eq. (12).  Therefore, $\lambda$ can be computed
explicitly when $\fmin$ is.  For example, this is possible in the
classical Bondi problem (also in presence of electron scattering, as
shown in Sect. 3), but in general this is impossible in presence of
the galaxy potential.  For this reason in order to determine the value
of $\lambda$ we must resort to numerical evaluation of $\fmin$. It is
then useful to be able to determine analytically the first correction
terms to the value of $\lambda$ due to the presence of the galaxy.  In
the following we present the general procedure for such determination.

The first step is to cast the function $f(x)$ of the problem under
scrutiny as the sum of two terms: one ($f_0$) so that the position of
the minimum, $\xz$, can be explicitly determined, $f_0'(\xz)=0$.  The
second term ($f_1$) is the perturbation term, depending on some small
ordering parameter $\epsilon\to 0$.  In practice, we suppose we can
write
\begin{equation} 
	f(x) = f_0(x) + f_1(\epsilon,x),
\end{equation}
with $ f_1(0, x)=0$.  If all the required regularity conditions are
satisfied, the position of the new minimum, $\xmin(\epsilon)$, is
given by
\begin{equation}
f'[\xmin (\epsilon)]=0,
\end{equation}
with 
\begin{equation}
\xmin (\epsilon) = \xz + \epsilon x_1+ \epsilon^2 x_2 + ... 
\end{equation}
By expansion of eq. (A2) and order balance one can in principle
determine the perturbation coefficients in eq. (A3) at the desidered
order. For example, provided that $f''_0(\xz) >0$,
it can be proved that 
\begin{equation}
x_1 = -{1\over f''_0(\xz)}\left[{\partial^2 f_1\over\partial\epsilon\partial 
    x}\right]_{(\epsilon,x)=(0,\xz)},
\end{equation}
and in such cases when $x_1 =0$
\begin{equation}
x_2 = -{1\over 2 f''_0(\xz) }\left[{\partial^3 f_1\over\partial\epsilon^2\partial 
    x}\right]_{(\epsilon,x)=(0,\xz)}. 
\end{equation}
Once the coefficients are determined, $\fmin$ is obtained by expansion
of eq. (A1): for example, in case of a regular perturbation $f_1$, it
is easy to show that at the first order
\begin{equation}
\fmin\sim f_0(\xz) + \epsilon \left[ {\partial
    f_1(\epsilon,\xz)\over\partial\epsilon}\right]_{\epsilon=0}.
\end{equation}
In Sect. 5 simple perturbation cases are described for the case of an
Hernquist galaxy.  One is obtained for $\xi\to\infty$, so that
$\epsilon =1/\xi$, the unperturbed problem is the Bondi accretion
(with or without electron scattering), with $f_0$ given in eq. (26),
and $\xz$ given by eq. (27), while $f_1=\calR\epsilon
x^{4(\gamma-1)\over\gamma+1} /(1+\epsilon x)$.  A second, strictly
related case, is that of $\epsilon=\calR\to 0$ so that the unperturbed
problem is the same as in the previous case, but now $f_1=\epsilon
x^{4(\gamma-1)\over\gamma+1}/(\xi+x)$. Finally, we consider
$\epsilon=\xi\to 0$. In this case the galaxy core radius vanishes, so
that the unperturbed problem is obtained by considering the Bondi
accretion on a MBH of mass $(\chi+\calR)\Mbh$, with unpertburbed
solution $\xz$ given by eq. (48), while $f_1=-\calR\epsilon
x^{4(\gamma-1)\over\gamma+1}/(\epsilon + x)$.

\section{Adiabatic accretion in general potential}
\label{app:adia}

Here we show that for $\gamma=5/3$ the function $f'$ relative to Bondi
accretion with electron scattering and in presence of a galaxy, is
definite positive for $x \ge 0$, so that $f$ is monotonically
increasing and the (only) minimum over the physical domain is attained
for $x \to 0$.  In fact, for a spherical galaxy model of density
profile $\rhog$, finite total mass $\Mg$, and with a central MBH of
mass $\Mbh$, the total potential in presence of electron scattering
can be written as (see eq. 39 and Binney \& Tremaine 1987)
\begin{equation} 
\phit (r)=-G {\Mg(r) +\chi\Mbh\over r} - 4 \pi G \int_r^{\infty} \rhog(r') r'\, dr'.
\end{equation}
With the introduction of the dimensionless total potential $\psi_{\rm
  t}\equiv -\rb\phit/(G \Mbh)$, the general form of eq. (47) reads
\begin{equation}
f= x \left({3\over 2} + \psi_{\rm t} \right),
\end{equation}
so that 
\begin{equation}
f'= {3\over 2} + 4\pi \int_x^{\infty}  \rhotil_{\rm g}(x') x'\, dx' > 0,
\end{equation}
where $\rhotil_{\rm g}=\rb^3 \rhog/\Mbh$, and this concludes the
proof.  It follows that for model galaxies with $\lim_{r\to
  0}r\phig(r)=0$, for $\gamma=5/3$ one has $\fmin=\chi$ so that
$\lambdat=\lambdaes=\chi^2/4$. The Hernquist model with $\xi>0$
satisfies this hypothesis, but not when $\xi=0$. This is the reason
behind the fact that the $\gamma=5/3$ case for the Hernquist model is
singular, in the sense that the properties of the $\xi=0$ case cannot
be obtained as the limit for $\xi\to 0$, as mentioned in Sect. 5.1.

Note that for $\gamma<5/3$ eq. (47) shows that the exponent of $x$ in
eq. (B2) is $< 1$, and the procedure above now confirms that $f'$ can
change sign, depending on the specific shape of $\phit$.

\section{Analytical properties of Hernquist case} 
\label{app:Hernquist}

In the general case of Bondi accretion with electron scattering in a
Hernquist galaxy with a central MBH and $\xi=\rg/\rb>0$ (the case
$\xi=0$ is trivial, as discussed in Sect. \ref{sec:Hernquistgal}), the
critical points of $f$ are placed at the zeroes of
\begin{equation}
f'={4x^{-2(3-\gamma)\over\gamma+1}g(x)\over (\gamma+1)(\xi+x)^2},
\end{equation}
where
\begin{equation} 
\begin{split}
&\ g=x^3 - {(5-3\gamma)(\chi+\calR)-8\xi\over 4} x^2 + \\
&\ {\xi\left[2(\gamma-1)\calR + 2\xi - (5-3\gamma)\chi \right]\over 2}
x -{\xi^2(5-3\gamma)\chi\over 4},
\end{split}
\end{equation}
as obtained from eq. (47), and $\chi =1-L/\Ledd$.  In the isothermal
and in the adiabatic monoatomic cases this reduces to
\begin{equation}
g=
\begin{cases}
\displaystyle{x^3-{\chi+\calR-4\xi\over 2} x^2+\xi(\xi-\chi)x
  -{\xi^2\chi\over 2},}\\
\displaystyle{x \left[ x^2 +2\xi x + \xi \left( \xi + {2\calR\over 3} \right)
\right],}
\end{cases}
\end{equation}
respectively, and in the last case the positivity for $x>0$ is
apparent, in accordance with the general result of Appendix
\ref{app:adia}.  Note that in the limiting case of $\chi=0$, i.e. when
formally $L=\Ledd$, the problem reduces to Bondi accretion (without
electron scattering) in the galaxy potential well only: therefore the
obtained formulae can used also to discuss this special class of
solutions.

For $1 \le \gamma < 5/3$, $\xi>0$ and $\chi>0$, the constant term in
eq. (C2) is negative, while the cubic term is positive.  This means
that $f$ presents always at least one minimum at $x>0$.  However, it
may happens that for specific values of the parameters there are three
positive zeros of $f$, corresponding, for increasing $x$, to a
minimum, a maximum, and a minimum of $f$, respectively. This is
relevant not only for the theoretical implications but also for
numerical investigations. In fact, the position of the absolute
minimum of $f$ is usually obtained by numerical analysis, and it is of
great help to know whether there just one minimum or it is necessary
to investigate what of the minima is the absolute one, as required in
order to compute the critical accretion parameter $\lambda$.

For assigned values of the parameters the existence and the position
of the zeros of $f'$ can be determined by using the Sturm method or
from the theory of cubic equations.  In fact, after standard reduction
of eq.~(C2) to canonical form $z^3+pz+q$, one consider the sign of the
quantity $R=q^2/4+p^3/27$.  When $R>0$ the original equation has one
real (and so positive) root and two complex conjugate roots. For
$R<0$, there are three real roots, and for $R=0$ there are a real root
and a double (and so real) root. In the case of three real roots,
their positivity can be determined from the Descartes' sign rule.

The function $R$ associated with eq.~(C2) is the product of the factor
$\xi^2\calR$ times a cubic polynomial that can be ordered as a
function of $\xi$ or $\calR$ as:
\begin{equation} 
\begin{split}
 &\  64(\gamma+1)\xi^3 + 4(P_3\calR+12P_6)\xi^2 - \\
 &\  4 \left[ P_1\calR^2 + P_4\calR -3(5-3\gamma)\chi P_6 \right]\xi - \\
 &\ (5-3\gamma)^2(\chi+\calR)^2 \left[ 4(\gamma-1)^2\calR -P_6 \right]=\\
 &\ - 4(5-3\gamma)^2(\gamma-1)\calR^3 - (4P_1\xi+P_2) \calR^2 + \\
 &\ 2(P_3\xi^2 - 2P_4\xi -P_5)\calR + (\gamma+1)\left[4\xi+(5-3\gamma)
   \chi \right]^3,
\end{split}
\end{equation}
where the coefficients are
\begin{equation} 
\begin{cases}
P_1=2(\gamma-1)(\gamma^2-30\gamma+33), \\
P_2=(5-3\gamma)^2(11\gamma^2-18\gamma+3)\chi, \\
P_3=2(23\gamma^2+30\gamma-57), \\
P_4=(5-3\gamma)(11\gamma^2-6\gamma+15)\chi, \\
P_5=(5-3\gamma)^2(5\gamma^2-6\gamma-3)\chi^2, \\
P_6=(\gamma+1)(5-3\gamma)\chi.
\end{cases}
\end{equation}
For assigned values of $\calR, \xi, \gamma$, and $\chi$, is then
straightforward to determine numerically the number of real zeros of
$h$ by using eqs~(C4)-(C5), and successively their positivity checking
the number of sign variations in the coefficients of eq.~(C2). In the
next Sections we will use eq.~(C4) to deduce some analytical property
of the critical points in some limit case.

\subsection{Dependence on the galaxy core radius}

In terms of $\xi$, the leading term of eq.~(C4) is positive, so that
$R>0$ for fixed $\calR$ and sufficiently large values of $\xi$, and
there is only one minimum for $f$, as expected from physical
considerations.  The displacement of the minimum position with respect
to eq.~(27) can be determined by using the method in Appendix A.  For
very small values of $\xi$ and $5-3\gamma>0$ instead the problem
presents three real solutions for
$\calR>(\gamma+1)(5-3\gamma)\chi/4(\gamma-1)^2$: a careful application
of Descartes' rule shows that all the three zeros are positive.  In
fact, reducing $\xi$ at fixed $\calR$, the function $f$ flattens in
the outer regions, until $R=0$ and an horizontal flex appears, while
the only minimum is placed near the position $\xmin$ in eq.~(48):
again, the correttive terms can be obtained from Appendix A.  Reducing
further $\xi$, the double solution splits, and the
minimum-maximum-minimum structure appears.  The outer minimum deepens
until it becomes the absolute minimum, approaching $\xmin$ given in
eq.~(48), while the other two zeroes merge at $x=0$. With some work it
is possible to obtain the first terms of the asymptotic expansion for
$\xi\to 0$ of the positions of the three critical points of eq.~(C2),
ordered for increasing distance from the origin:
\begin{equation}
\begin{cases}
\displaystyle{x_1 \sim 
{2\calR (\gamma -1)-\chi (5-3\gamma)-\sqrt{\Delta}
\over (\chi+\calR) (5-3\gamma)}\xi ,}
\\
\displaystyle{x_2 \sim 
{2\calR (\gamma -1)-\chi (5-3\gamma)+\sqrt{\Delta}
\over (\chi+\calR) (5-3\gamma)}\xi ,}
\\
\displaystyle{x_3 \sim {(\chi+\calR) (5-3\gamma)\over 4} - {2\calR
    (3-\gamma)\xi\over (5-3\gamma)(\chi +\calR)},}
\end{cases}
\end{equation}
where $\Delta =\calR \left[ 4(\gamma-1)^2\calR -P_6 \right]$.  It is
apparent how the reality condition for $x_1$ and $x_2$ the first two
zeroes matches the reality condition obtained from eq.~(C4). Moreover,
again with some work from eq.~(C4), it can be proved that the reality
condition of the three zeroes for $\gamma\to 5/3$ and small $\xi$
requires that $\xi=O(5-3\gamma)^2$, so that the three critical points
in eq.~(C6) collapse into the origin for $\gamma\to5/3$, consistently
with the general result of Appendix B for adiabatic accretion.

\subsection{Dependence on the galaxy mass}

A different situation is obtained when considering eq. (C4) as a
function of $\calR$.  It is apparent how $R>0$ for $\calR \to 0$, and
there is only one real solution, as expected, because the problem
reduces to the classical Bondi problem.  The perturbed position of the
minimum can be determined again from Appendix A, with $x_0$ given in
eq.~(27).  Instead, $R<0$ for very large values of $\calR$ and
$\gamma>1$, and so eq.~(C2) admits three real solutions.  Descartes'
sign rule shows that the three zeros are positive. By using the
order-balance technique it can be shown that their asymptotic leading
term at fixed $\gamma$ and $\xi$ (in order of increasing distance from
the center) is given by
\begin{equation}
\ x_1 \sim {\chi\xi(5-3\gamma)\over 4(\gamma-1)\calR},
\ x_2 \sim {4\xi(\gamma-1)\over 5-3\gamma}, 
\ x_3 \sim {(5-3\gamma)\calR\over 4}. 
\end{equation}
One could be worried by the possibility that for $\gamma\to 5/3$, the
solution $x_2$ becomes larger than $x_3$: however it can be shown that
once the reality condition is verified this cannot happens, because in
order to have $R<0$, the mass ratio $\calR$ must diverge faster than
$(5-3\gamma)^{-2}$ for $\gamma\to 5/3$. The $\gamma=1$ case is
discussed in the following Section.

\subsection{The isothermal case}

A particularly simple case is the isothermal one.  In fact, for $\gamma=1$
eq. (C4) leads to the study of the sign of the quadratic polynomial
$\chi \calR^2+(2\chi^2-10\chi \xi-\xi^2)\calR +(\chi+2\xi)^3$.
Therefore, at variance with the $\gamma>1$ cases, in isothermal accretion
$R>0$ for large values 
of $\calR$ and fixed $\xi$, and only one (positive) minimum exists,
with the sonic point given by  the last identity in eq.~(C7) with $\gamma=1$.
In addition, $R>0$ for large values of $\xi$ and fixed $\calR$, with
the sonic point placed at $\xmin$ in eq.~ (27) with $\gamma=1$.
Finally $R>0$ independently of $\calR$ when
$\xi<4\chi$, and so again there is only one (positive) minimum.
For $\xi \ge 4\chi$ there are two positive values of the galaxy-to-MBH
mass ratio
\begin{equation}
\calR_{\rm max,min} ={-2\chi^2+10\chi \xi +\xi^2 \pm
  \sqrt{\xi(\xi-4\chi)^3}/\over 2\chi},
\end{equation}
so that $R<0$ for $\calR_{\rm min}<\calR<\calR_{\rm max}$: Descartes'
sign rule shows that the three zeros are all placed at $x>0$. In
particular, for $\xi \to \infty$, $\calR_{\rm min} \sim 8\xi$ and
$\calR_{\rm max} \sim \xi^2/\chi$.  Summarizing, in the isothermal
case, for $\xi < 4\chi$ there is only one minimum $\forall \calR>0$,
for $\calR < 27\chi$ there is only one minimum $\forall \xi >0$, and
there is only one minimum for large values of $\calR$ ($\xi$) and
fixed $\xi$ ($\calR$). The situation is clearly illustrated in
Fig. C1.

\begin{figure}
       \subfloat{\includegraphics[height=0.55\textwidth, width=0.49\textwidth]{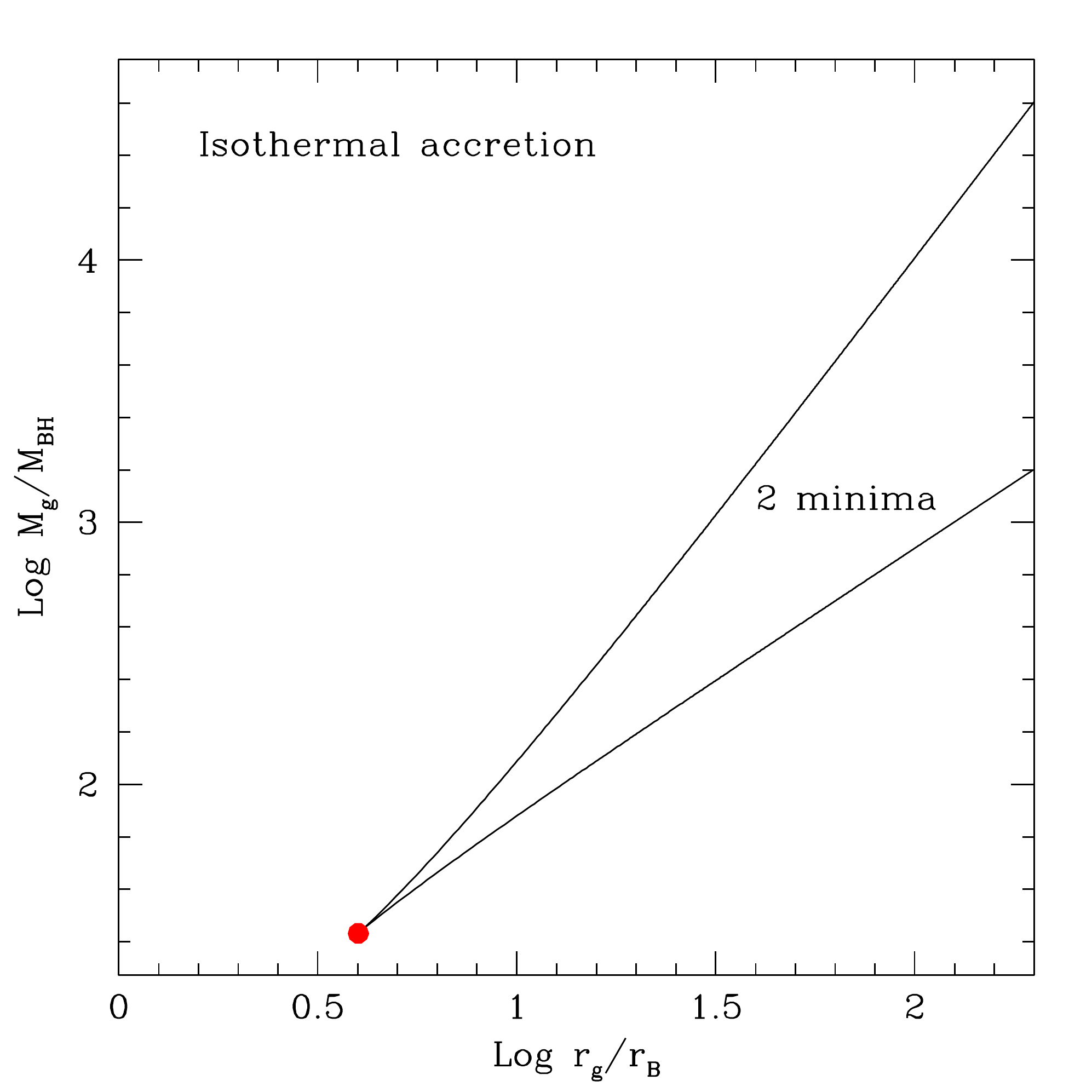}} \hfill
       \caption{Parameter space for the existence of minima relative
         to the isothermal Bondi accretion in a Hernquist galaxy. For
         simplicity the effect of electron scattering is not taken
         into account, i.e., we assume $\chi=1$ in eq. (C2).  The two
         solid lines show the boundaries given by eq. (C7): in the
         region between the two lines the problem presents three
         positive zeroes, corresponding to two minima and one
         maximum. Outside the infinite triagular regions only one
         minimum exists. The coordinate of the red dot are $(\xi,
         \calR)=(4,27)$. Note that for the fiducial values $(100,
         1000)$ the accretion presents two minima.}
      \label{fig:isodelta}
\end{figure}

\section*{Acknowledgements}
L.C. is grateful to G. Bertin and J. Ostriker for useful discussions.
L.C. and S.P. were supported by the MIUR grant PRIN 2010-2011, project
``The Chemical and Dynamical Evolution of the Milky Way and Local Group
Galaxies'', prot. 2010LY5N2T.

\end{document}